\documentclass[aps,pre,showkeys,reprint,twoside]{revtex4-1}
\pdfoutput=1

\usepackage[utf8]{inputenc}
\usepackage[T1]{fontenc}
\usepackage{lmodern} 
\usepackage{textcomp}

\usepackage{amsmath,mathtools}
\usepackage{amssymb}
\usepackage{amsthm}
\usepackage{mathrsfs}
\usepackage{accents}
\usepackage{bm}

\usepackage{sistyle}

\usepackage{graphicx}
\usepackage[dvipsnames]{xcolor}
\usepackage{ragged2e}
\usepackage[boxed]{algorithm2e}

\usepackage{fancyhdr}

\usepackage[pdfborder={0 0 0.4}]{hyperref} 

\let\oldparagraph\paragraph
\renewcommand{\paragraph}[1]{\oldparagraph{#1}\phantomsection}

\SetAlgorithmName{ALGO.}{Algo.}{List of algorithms}
\SetAlgoCaptionSeparator{.\,}
\SetAlCapSkip{1ex}
\SetAlCapSty{textnormal}

\SetAlgoCaptionLayout{myAlgoCaptionLayout}
\SetKwInput{KwFunc}{Function}

\makeatletter
\def\p@paragraph      {\thesection:}
\def\p@subparagraph   {\thesection:\theparagraph:}
\makeatother

\newcommand{\mytitle}{Addressing the gas kinetics Boltzmann equation with branching-path statistics}
\SIdecimalsign{\pnt}
\SIunitdot{\,}

\SetSymbolFont{largesymbols}{bold}{OMX}{txex}{b}{n}

\newlength{\Graphwidth}
\newlength{\graphwidth}
\newlength{\dhatheight}
\newcommand{\doublehat}[1]{%
    \settoheight{\dhatheight}{\ensuremath{\hat{#1}}}%
    \addtolength{\dhatheight}{-0.25ex}%
    \hat{\vphantom{\rule{1pt}{\dhatheight}}%
    \smash{\hat{#1}}}
}
\newlength{\dbarheight}
\newcommand{\doublebar}[1]{%
	{\mathchoice%
	{	\settoheight{\dbarheight}{\ensuremath{\displaystyle\bar{#1}}}%
		\addtolength{\dbarheight}{-0.2ex}%
		\bar{\vphantom{\rule{1pt}{\dbarheight}}%
		\smash{\bar{#1}}}}%
	{	\settoheight{\dbarheight}{\ensuremath{\textstyle\bar{#1}}}%
		\addtolength{\dbarheight}{-0.2ex}%
		\bar{\vphantom{\rule{1pt}{\dbarheight}}%
		\smash{\bar{#1}}}}%
	{	\settoheight{\dbarheight}{\ensuremath{\scriptstyle\bar{#1}}}%
		\addtolength{\dbarheight}{-0.14ex}%
		\bar{\vphantom{\rule{1pt}{\dbarheight}}%
		\smash{\bar{#1}}}}%
	{	\settoheight{\dbarheight}{\ensuremath{\scriptscriptstyle\bar{#1}}}%
		\addtolength{\dbarheight}{-0.08ex}%
		\bar{\vphantom{\rule{1pt}{\dbarheight}}%
		\smash{\bar{#1}}}}%
}}

\newcommand{\texte}[1]{{\text{#1}}}
\newcommand{\textc}[1]{{\mathit{#1}}}
\newcommand{\indic}[1]{{\textnormal{#1}}}
\newcommand{\sfo}{\rule{0pt}{1ex}}

\newcommand{\fnl}{a}
\newcommand{\gnl}{b}
\newcommand{\xx}{{\vec{x}}}

\newcommand{\XX}{{\vec{X}}}
\newcommand{\YY}{{\vec{Y}}}

\newcommand{\ud}{d}
\DeclareMathOperator{\id}{id}
\newcommand{\pdf}[1]{{p_{#1}}}

\newcommand{\esmbb}[1]{\mathbb{#1}}

\newcommand{\doms}[1]{\mathcal{#1}}

\newcommand{\intersep}{\mspace{-5mu}:\mspace{-2mu}}
\DeclareMathOperator{\esp}{E}
\DeclareMathOperator{\var}{Var}
\DeclareMathOperator{\cov}{Cov}
\newcommand{\est}[1]{\tilde{#1}}
\newcommand{\grd}{q}

\newcommand{\vgrd}{h}
\newcommand{\wgrd}{w}
\newcommand{\dgrd}{\doms{D}}

\newcommand{\reels}{\esmbb{R}}
\newcommand{\ntrs}{\esmbb{N}}
\newcommand{\rec}{\textc{rec}}

\newcommand{\ff}{f}
\newcommand{\FF}{\est{F}}
\newcommand{\Ff}{\est{f}}

\newcommand{\seceff}{\sigma}
\newcommand{\sedf}{{\seceff\mspace{-2mu}_\indic{F}}}

\newcommand{\rr}{\vec{r}}
\newcommand{\RR}{\vec{R}}
\newcommand{\cc}{\vec{c}}
\newcommand{\CC}{\vec{C}}
\newcommand{\ccd}{\cc^{\:2}}
\newcommand{\ccp}{{\cc^{\mspace{3mu}\prime}}}
\newcommand{\CCP}{{\CC'}}
\newcommand{\cce}{{\cc_*}}
\newcommand{\CCE}{{\CC_*}}
\newcommand{\ccpe}{{\cc^{\mspace{3mu}\prime}_*}}
\newcommand{\CCPE}{{\CC'_*}}
\newcommand{\ccrn}{g}
\newcommand{\ccr}{{\vec{\ccrn}}}
\newcommand{\uu}{\vec{u}}
\newcommand{\UU}{\vec{U}}
\newcommand{\uup}{{\uu^{\mspace{1.5mu}\prime}}}
\newcommand{\UUP}{{\UU'}}
\newcommand{\acc}{\vec{a}}
\newcommand{\spcc}{{E_c}}
\newcommand{\spcu}{{E_u}}
\newcommand{\spcr}{{E_r}}
\newcommand{\vnul}{\vec{0}}
\newcommand{\vnabla}{\vec{\nabla}}
\newcommand{\ux}{\vec{u\mspace{-1mu}_x}}

\newcommand{\opcolb}{{C\mspace{-1mu}_\indic{B}}}
\newcommand{\Frac}{\textc{Fra\mspace{-2.5mu}c}\,}
\newcommand{\cmol}{\eta}
\newcommand{\Mvv}{\vec{v}}
\newcommand{\Mvvp}{{\Mvv\mspace{2mu}'}}
\newcommand{\cqm}{{c_\indic{q}}}
\newcommand{\bkwk}{K}
\newcommand{\raid}{\omega}
\newcommand{\cqmeq}{{c_\indic{q,\,eq}}}
\newcommand{\rcqm}{\epsilon}
\newcommand{\phas}{\phi}
\newcommand{\phaz}{{\phi_0}}
\newcommand{\cok}{\indic{cok}}
\newcommand{\cqmcok}{{c_\indic{q,\,cok}}}
\newcommand{\Bedf}{{B_\indic{F}}}
\newcommand{\eq}{\indic{eq}}
\newcommand{\final}{\indic{f}}

\newcommand{\sis}{{s_\indic{is}}}

\newcommand{\ft}{f}
\newcommand{\kc}{\hat{\nu}}
\newcommand{\FT}{\est{F}}
\newcommand{\Ft}{\est{f}}

\newcommand{\matid}{\doublebar{1}}

\newcommand{\rrb}{{\rr_b}}
\newcommand{\ccb}{{\cc_b}}

\newcommand{\vecb}{\vec{b}}
\newcommand{\frec}{\hat{\nu}}
\newcommand{\fret}{{\nu_\indic{t}}}
\newcommand{\frecc}{\doublehat{\nu}}

\newcommand{\algobs}{\rhd}
\newcommand{\algo}{\mathchoice
	{\protect\raisebox{0.25ex}{$\displaystyle \mspace{1mu}\algobs$}}
	{\protect\raisebox{0.25ex}{$\textstyle \mspace{1mu}\algobs$}}
	{\protect\raisebox{0.15ex}{$\scriptstyle \mspace{1mu}\algobs$}}
	{\protect\raisebox{0.1ex}{$\scriptscriptstyle \mspace{1mu}\algobs$}}
}

\begin{document}

\title{\mytitle}
\author{Guillaume Terr\'ee}
\email[Corresponding author: ]{gterree@mines-albi.fr}
\author{Mouna \surname{El Hafi}}
\affiliation{Centre RAPSODEE, UMR CNRS 5302, IMT Mines Albi, Université de Toulouse, F-81013 Albi CT, France}
\author{St\'ephane Blanco}
\author{Richard Fournier}
\affiliation{UPS, CNRS, INPT, LAPLACE UMR CNRS 5213, Université de Toulouse, 118 route de Narbonne, F-31065 Toulouse, Cedex 9, France}
\author{J\'er\'emi Dauchet}
\affiliation{Université Clermont Auvergne, Clermont Auvergne INP, CNRS, Institut Pascal, F-63000 Clermont-Ferrand, France}
\author{Jacques Gautrais}
\affiliation{Centre de Recherches sur la Cognition Animale (CRCA), Centre de Biologie Intégrative (CBI), Université de Toulouse;
CNRS, UPS, F-31000 Toulouse, France}
\date{\today}
\begin{abstract}
This article proposes a statistical numerical method to address gas kinetics problems obeying the Boltzmann equation. This method is inspired by Monte-Carlo algorithms used in linear transport physics, where virtual particles are followed backwards in time along their paths. The non-linear character of gas kinetics translates, in the numerical simulations presented here, into branchings of the virtual particle paths. The obtained algorithms have displayed in the few tests presented here two noticeable qualities: (1) they involve no mesh, and (2) they allow one to easily compute the gas density at rarefied places of the phase space, for example at high kinetic energy.
\end{abstract}
\newcommand{\mykeywords}{Monte-Carlo method; branching random process; integral formulation; unbiased estimator; non-linear Boltzmann equation; gas kinetics}
\maketitle
\hypersetup{pdfauthor={Guillaume Terr\'ee, Mouna \surname{El Hafi}, St\'ephane Blanco, Richard Fournier, J\'er\'emi Dauchet, and Jacques Gautrais}, pdftitle={\mytitle}, pdfkeywords={\mykeywords}}

\section{Introduction}
\label{sec:intro}

We appraise the possibility of using the Monte-Carlo Method (MCM) to tackle the non-linear Boltzmann model for gas dynamics. 
In paragraph~\ref{sec:intro:boltzmann-equation} below, we recall the corresponding equation.
Then, in Par.~\ref{sec:intro-proprietes-MMC}, the design of the classical DSMC method to solve it is briefly presented.
In Par.~\ref{sec:intro:points-de-vue-MMC}, we present the relevance of MCM as a tool to tackle linear transport models, and the reasons why to try to extend it to non-linear cases.
In Par.~\ref{sec:intro-NLMC-history}, we list some successful improvements that have been made so far towards such an extension.
Finally, in Par.~\ref{sec:intro:our-proposal}, we present how that recent progress can be combined to translate the Boltzmann model into a MCM practice.

\paragraph{We aim to solve the Boltzmann equation}\label{sec:intro:boltzmann-equation} for mono-atomic gas in rarefied regimes. This equation reads \cite{Cercignani1988}
\begin{subequations}
\label{eq:boltzmann}
\begin{equation}
D (\ff) = \opcolb (\ff),
\end{equation}
where $\ff$ is the distribution function of molecules at time $t$ over positions $\rr$ and velocities $\cc$, $D$ is the particle derivative
\begin{multline}
D (\ff) (\rr;\cc;t) = \sfo \\
 \partial_t \ff (\rr;\cc;t) + \cc \cdot \vnabla_{\!\rr} \,\ff (\rr;\cc;t) + \vnabla_{\!\cc} \cdot [ \ff (\rr;\cc;t) \acc (\rr;\cc;t) ],
\end{multline}
$\acc$ is the acceleration vector of molecules due to long-range forces, $\opcolb$ is the Boltzmann collision operator
\begin{multline}
\label{eq:boltzmann-opcol}
\opcolb (\ff) (\rr;\cc;t) = \int_\spcc \ud\cce \int_\spcu \ud\uup \,\ccrn \,\sedf ( \ccrn; \uup \cdot \uu_\ccr ) \times \sfo \\
[ \ff (\rr;\ccp;t) \ff (\rr;\ccpe;t) - \ff (\rr;\cc;t) \ff (\rr;\cce;t) ],
\end{multline}
$\spcc$ and $\spcu$ denote respectively the spaces of velocities and directions (so $\spcu$ is the unit sphere of $\spcc$), $\cce$ is the velocity of the collision partner of a molecule at $(\rr;\cc;t)$, $\uup$ is the incoming direction of this molecule in the collision barycentric frame,
\begin{equation}
\label{eq:speeds-sphere}
\left\{\begin{aligned}
\ccr &= \cc - \cce \\
\ccrn &= \| \ccr \| \\
\uu_\ccr &= \ccr / \ccrn \\
\ccp &= \tfrac{1}{2} (\cc + \cce + \ccrn \uup) \\
\ccpe &= \tfrac{1}{2} (\cc + \cce - \ccrn \uup)
\end{aligned}\right.,
\end{equation}
\end{subequations}
i.e., $\uu_\ccr$ is the exiting direction of the molecule in the collision barycentric frame and $\ccrn$ is the relative speed of colliding molecules, and $\sedf$ is the differential collision cross-section.

The Boltzmann equation~\eqref{eq:boltzmann} is non-linear due to the collision operator $\opcolb$.

\begin{figure*}
\parbox{\Graphwidth}{\parbox{\graphwidth}{\includegraphics{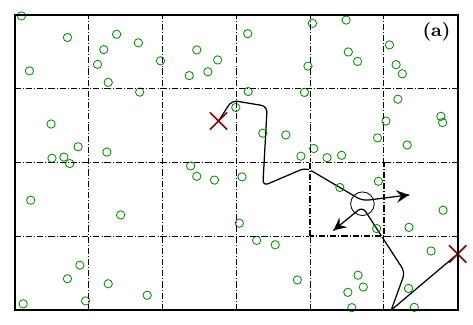}}
\hfill
\parbox{\graphwidth}{\includegraphics{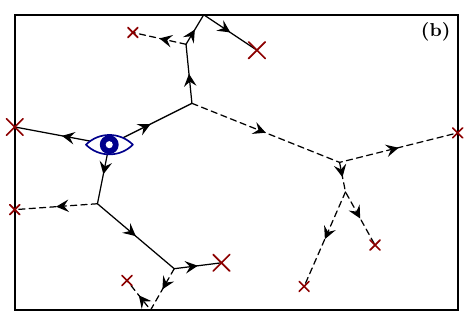}}}
\caption{
Illustration of two numerical methods in gas kinetics, for computing the molecule density at a given time and at a given point inside a rectangular domain: (a) the methods of the Direct Simulation Monte-Carlo family, and (b) the method presented in the present article.\\
(a) The DSMC methods are designed as analogies of the gas kinetics. Virtual molecules (green circles) are introduced at the initial condition or at some boundaries (which is illustrated on two molecules with red crosses). Then, trajectories of molecules are built by iterating a two-step process of advection and collision (two molecule trajectories are displayed with broken black arrows). To implement collisions between pairs of molecules, all molecules histories are built simultaneously. Since the virtual molecules never cross exactly, a proximity criterion is introduced (black circle in a highlighted cell of a grid) to identify potential collision partners, and the collisions are made to occur with suitable probability.
The virtual molecules must be numerous enough for their swarm to statistically represent the density field, for computing accurately the collision statistics. This method requires one to solve the evolution of the entire field from initial condition up to the final time of interest.\\
(b) The method we present starts from the location and final time of interest (probe point, eye).
From there, molecules' paths (plain lines) are sampled backwards in time (as shown by the way of arrows) up to the initial condition or boundaries (red crosses). 
Collision events are sampled along these paths (forking dashed lines).
The collision partner properties needed to resolve the collision output being unknown, they are themselves sampled at the collision points using the same algorithm as for the probe point.
Since these potential collision partners can themselves encounter collisions, the algorithm is branching.
In this way, the branching algorithm builds particle trees, where the probe point is the root and the red crosses are the leaves.
}%
\label{fig:differents-suivis}
\end{figure*}

\paragraph{A usual way of solving}\label{sec:intro-proprietes-MMC} the Boltzmann equation~\eqref{eq:boltzmann} is the Direct Simulation Monte Carlo (DSMC) method \cite{Bird1994, Bird1998, Oran1998, Murrone2011}. The numerical schemes of the DSMC family [illustrated in Fig.~\ref{fig:differents-suivis}(a)] share the following operating principles: they build the story of a swarm of statistically representative molecules, and because these cannot cross exactly, collisions are introduced using a proximity criterion \cite{Bird1994, Bird1998, Oran1998, Murrone2011}. Many variants are used today, depending, for example, on the collision quadrature used in time and space \cite{Khisamutdinov2004, Stefanov2011}, or on whether an Eulerian fluid model is used together with the DSMC formulation \cite{Homolle2007, Degond2010}.
In the DSMC method, the non-linearity of the collision operator is handled with discretization of time and space, which underpins the proximity criteria which pilot the choice of molecules to make collide.

\paragraph{As an alternative,}\label{sec:intro:points-de-vue-MMC} we present here the possibility of extending the Monte-Carlo method classically used in linear transport physics (such as radiative transfer) \cite{Howell1998, Metropolis1949}.

For linear transport models, the Monte-Carlo method basically consists in simulating, with the help of a random number generator, the history of numerous independent particles, from which average quantities can be computed \cite{Howell1998}. Such an algorithm can be perceived as an analogous computation of the physical phenomenon under study \cite{Howell1998}. Refined algorithms have diversified from the raw mimicry of the physical model, for example by the use of tailored sampling with compensating weighting in order to obtain a better computational efficiency \cite{Spanier1970, DelaTorre2014}.

Alternatively, the MCM can be interpreted as a statistical quadrature for computing high-dimensional integrals \cite{Metropolis1949, Dimov2008,Dunn2012}. If a quantity $\grd$ is expressed as an integral over some space $\dgrd$ of a function $\vgrd$, i.e., 
\begin{subequations}
\label{eq:intro-MMC-principle}
\begin{align}
\grd &= \int_\dgrd \ud\xx \, \vgrd (\xx),
\intertext{a Random Variable (RV) $\XX$ can be introduced, with probability density $\pdf{\XX}$ such as $\forall \xx \in \dgrd, \vgrd (\xx) \neq 0 \Rightarrow \pdf{\XX} (\xx) \neq 0$, and then with $\wgrd (\xx) = \vgrd (\xx) \big/ \pdf{\XX} (\xx)$ and $\esp$ the statistical expectation,}
\grd &= \int_\dgrd \pdf{\XX} (\xx) \ud\xx \, \wgrd (\xx) \\
 &= \esp [ \wgrd (\XX) ].
\end{align}
\end{subequations}
So, in order to compute $\grd$, a large sample of $\XX$ is drawn, yielding as many realizations of the RV $\wgrd (\XX)$, the sample mean of which is an estimator of the expectation $\grd$. This estimator comes with its confidence intervals, deduced from the variance of the sample according to the central limit theorem. All the work presented here stems from this integral interpretation.

In linear transport physics, the MCM has come to be widely used due to its many qualities.
It allows an efficient management of multi-dimensional phase space and can take into account numerous physical phenomena in a single algorithm \cite{Dimov2008, Dunn2012}.
It allows the computation of a quantity in this phase space (picking a probe point of interest) at arbitrary time with no need to solve the evolution of the entire field since the initial condition.
For this, particle paths are sampled in a \emph{reversed time} direction up to a boundary or initial condition \cite{Andreucci1985, Gurov2002, Howell1998, Galtier2017a, Carter1975, Lu2005}.
The estimator displays a null systematic error compared to the mathematical and physical model, and comes with confidence intervals.
There is no need to discretize the space nor the time, so solving problems involving complex geometries yields no conceptual difference nor technical bottleneck \cite{Villefranque2019}.
It also allows one to calculate sensitivities from within the main simulation \cite{DeLataillade2002, Tregan2020}, and parallelization is straightforward.

Our proposal is to appraise the MCM  to compute $\ff (\rr;\cc;t)$ governed by Eq.~\eqref{eq:boltzmann}, while preserving these many qualities.

\paragraph{In non-linear transport physics,}\label{sec:intro-NLMC-history} the MCM presented above is still commonly believed to be unusable:
\begin{quote}
``So far as the author is aware, the extension of Monte Carlo methods to nonlinear processes has not yet been accomplished and may be impossible.'' \cite{Curtiss1954}

``Monte Carlo methods are not generally effective for nonlinear problems mainly because expectations are linear in character.'' \cite{Kalos2008}

``A nonlinear problem must usually be linearized in order to use Monte Carlo technique.'' \cite{Kalos2008}
\end{quote}
Indeed, the MCM is the computation of expectations, and any non-linearity seems to preclude the expression of the sought quantity as an expectation (as detailed in \cite{Dauchet2018}, and in Par.~\ref{sec:intro-trick} below).
However, several examples lead us to moderate these statements.

The first examples are in the frame of the work of Ermakov \emph{et al.\@ }\cite{Ermakov1976} using the MCM for solving non-linear kinetic equations. Gurov \cite{Gurov1992, Dimov2000} used branching estimation processes in the MCM to solve Fredholm equations of the second kind with polynomial non-linearities. Rasulov \emph{et al.\@ }\cite{Rasulov2004, Rasulov2010, Raimova2016} used the MCM to solve the inhomogeneous equation of heat with a source term depending non-linearly on the temperature. These techniques allowed the extension of the MCM towards a broader class of non-linear problems \cite{Labordere2014}.

Null Collision Algorithms (NCAs) are another known example of handling non-linearities in the MCM, although they are drawn from linear transport (e.g., radiative transfer physics \cite{Galtier2013}, image synthesis \cite{Novak2014, Szirmay-Kalos2017}, semiconductor physics \cite{Rees1968}, neutronics \cite{Woodcock1965, Spanier1966, Coleman1968}, and plasma physics \cite{Skullerud1968}). These algorithms are used there to handle the high variability of the extinction rate along the ballistic trajectories of particles (\emph{extinction} means here and throughout this article \emph{absorption} or \emph{outward scattering}). Their idea is to add virtual and ineffective colliders in order to make constant the total collision frequency. The NCAs are justified by the equivalence between the original and the modified transport problems \cite{Galtier2013, Novak2014, Szirmay-Kalos2017, Rees1968, Woodcock1965, Spanier1966, Coleman1968, Skullerud1968}; but they can also be viewed as a way to account for the non-linearity of the exponential Beer extinction law \cite{Longo2002, Georgiev2019}.

Drawing inspiration from the NCAs, our group recently adapted the MCM to several cases of non-linear physics with radiative transfer \cite{Dauchet2018}. Dauchet \cite{Dauchet2012} estimated the global productivity of a photobioreactor containing photosynthetic microalgae, despite the non-linear coupling between the local productivity at each point in the tank and the radiative transfer through the tank. Farges \cite{Farges2014} computed the productivity of a solar power plant over one year, accounting for a conversion efficiency depending on the instantaneous throughput. Dauchet \emph{et al.\@ }and Charon \emph{et al.\@ }\cite{Dauchet2015, Charon2015} evaluated the radiative properties of some microalgae using the electromagnetic theory, accounting for their size and shape distribution, despite the fact that radiative intensity is quadratic in the electric field amplitude for which they were having an estimator. These works are based on a technique which handles, in the MCM, multi-folded integrals with one non-linearity in between \cite{Dauchet2018}.

\paragraph{Combining this progress}\label{sec:intro:our-proposal} from \cite{Ermakov1976, Gurov1992, Dimov2000, Rasulov2004, Rasulov2010, Raimova2016, Labordere2014, Galtier2013, Novak2014, Szirmay-Kalos2017, Rees1968, Woodcock1965, Spanier1966, Coleman1968, Skullerud1968, Longo2002, Georgiev2019, Dauchet2018, Dauchet2012, Farges2014, Dauchet2015, Charon2015}, we develop below the theoretical framework to extend the MCM to the non-linear case of Eq.~\eqref{eq:boltzmann}, and offer some test-bed examples.
In the spirit of linear transport models, the main idea is to write algorithms to estimate $\ff$ at chosen probe points by collecting numerous histories of particles, sampling their walk and scattering events backwards in time. 
The distinct point in the present case, is that the collision operator conditioning those histories is not given by a field which is external to the process, so it must be itself estimated.
Namely, when tracking the history of a particle backwards in time, the density of collision partners being unknown, it is sampled by backtracking the histories of the partners every time it is needed.
The partners' histories being themselves conditioned by collision events with other partners, the estimation procedure of $\ff(\rr;\cc;t)$ is thus a \emph{branching process} \cite{Harris1964} [Fig.~\ref{fig:differents-suivis}(b)].  

To technically derive the corresponding algorithm, our methodology stems from the point of view on the MCM as a quadrature, as discussed in Par.~\ref{sec:intro:points-de-vue-MMC}.
First, the Boltzmann equation is translated into a purely integral counterpart, where $\ff(\rr;\cc;t)$ is expressed as an integral of $\ff$ at other locations in the past and of boundary and initial conditions.
Second, random variables are introduced for each integration variable, which leads to recursive Monte-Carlo algorithms to evaluate $\ff(\rr;\cc;t)$.

Being based on the same theoretical framework, the Monte-Carlo algorithms presented below preserve some interesting properties of the MCM mentioned above. 
Some of them are the following:
\begin{itemize}
\item There is no mesh nor discretization of space or time. Resolving problems involving complex geometries brings no conceptual difference nor technical blockage.  
\item The estimator has a null systematic error relative to the equation being solved. The only source of uncertainty is the statistical noise, which can be evaluated through the sample variance as mentioned in Par.~\ref{sec:intro:points-de-vue-MMC}.
\item The scarcity of molecules at the probe point does not compromise the relative accuracy of the estimation of the quantity of molecules there (at probe position and at probe velocity).
\end{itemize}



\paragraph{This article is organized as follows.} Section~\ref{sec:principle} shows how a Monte-Carlo algorithm being a branching estimation process can handle the non-linearity of the Boltzmann collision operator. 
We appraise the behavior and practicability of this algorithm in Sec.~\ref{sec:bkw}. 
The problem of sampling collisions, in a medium where the distribution of collision partners is unknown, is tackled in Sec.~\ref{sec:toy}, by using NCAs; this is illustrated and tested in the same section on a toy model. Sections~\ref{sec:ht} and \ref{sec:mix} are dedicated to the test of our proposal on two physical situations described by the Boltzmann equation~\eqref{eq:boltzmann} in its full complexity, one of which has an explicit solution to be compared with. 
We conclude in Sec.~\ref{sec:persp}.

\section{Principle of the method}
\label{sec:principle}

\paragraph{In the works \cite{Ermakov1976, Gurov1992, Dimov2000, Rasulov2004, Rasulov2010, Raimova2016, Labordere2014, Galtier2013, Novak2014, Szirmay-Kalos2017, Rees1968, Woodcock1965, Spanier1966, Coleman1968, Skullerud1968, Longo2002, Georgiev2019, Dauchet2018, Dauchet2012, Farges2014, Dauchet2015, Charon2015}}\label{sec:intro-trick} that we cited in Par.~\ref{sec:intro-NLMC-history}, which are about the MCM applied in non-linear physics, the non-linearity is not circumvented by using linearization, but by increasing the dimension of the sampling space.

As a theoretical illustration, let us consider now $\grd$ as the expectation 
\begin{subequations}
\begin{equation}
\grd = \esp [ \fnl (\XX) \wgrd (\XX) + \gnl (\XX) ],
\end{equation}
where 
$\fnl$ and $\gnl$ are two deterministic functions. 
Let us further suppose that $\wgrd (\XX)$ is itself an expectation of $\tilde{\wgrd} (\XX;\YY)$ over the RV $\YY$,
\begin{equation}
\wgrd (\XX) = \esp_{|\XX} [ \tilde{\wgrd} (\XX;\YY) ],
\end{equation}
where the subscript ${|\XX}$ means ``for a given value of $\XX$''.
We have
\begin{equation}
\grd = \esp \{ \fnl (\XX) \esp_{|\XX} [ \tilde{\wgrd} (\XX;\YY) ] + \gnl (\XX) \}.
\end{equation}
The statistical expectation is linear and projective, we also have
\begin{equation}
\grd = \esp [ \fnl (\XX) \tilde{\wgrd} (\XX;\YY) + \gnl (\XX) ].
\end{equation}
\end{subequations}
As a result, we do not need to know exactly $\wgrd (\XX)$ for each value of $\XX$ in order to evaluate $\grd$. Instead, $\grd$ can be estimated by first sampling $\XX$ and then $\YY$, namely sampling the RVs couple $(\XX;\YY)$. This is one of the principal advantages of the Monte-Carlo method: its complexity increases linearly (and not exponentially) with the dimension of the configuration space \cite{Metropolis1949}.

At first sight, this appears to be practicable only for a linear coupling law $\wgrd \to \grd$. 
Indeed, if we had
\begin{subequations}
\begin{equation}
\grd =  \esp [ \fnl (\XX) \wgrd (\XX)^2 ],
\end{equation}
then
\begin{equation}
\grd \neq \esp [ \fnl (\XX) \tilde{\wgrd} (\XX;\YY)^2 ],
\end{equation}
\end{subequations}
in the same way as $(\frac{u + v}{2})^2 \neq \frac{u^2 + v^2}{2}$.

However, we could consider that
\begin{multline}
\esp_{|\XX} [ \tilde{\wgrd} (\XX;\YY\!_1) \tilde{\wgrd} (\XX;\YY\!_2) ] =  \wgrd (\XX)^2 + \sfo \\ 
\cov_{|\XX} \bm( \tilde{\wgrd} (\XX;\YY\!_1); \tilde{\wgrd} (\XX;\YY\!_2) \bm),
\end{multline} 
where $\YY\!_1$ and $\YY\!_2$ are two RVs identically distributed as $\YY$, and $\cov$ denotes the covariance. If $\YY\!_1$ and $\YY\!_2$ are independent, this covariance is null.
Hence, considering two RVs $\YY\!_1$ and $\YY\!_2$ independent and identically distributed as $\YY$, we have
\begin{equation}
\grd = \esp [ \fnl (\XX) \wgrd (\XX)^2 ] = \esp [ \fnl (\XX) \tilde{\wgrd} (\XX;\YY\!_1) \tilde{\wgrd} (\XX;\YY\!_2) ].
\label{eq:espsquare}
\end{equation}
$\grd$ can be estimated by sampling the RVs triplet $(\XX;\YY\!_1;\YY\!_2)$ \cite{Dauchet2015, Charon2015}. 

The very same principle can be extended to monomial and polynomial functions \cite{Gurov1992, Dimov2000}, following
\begin{equation}
\forall n \in \ntrs, \esp [ \fnl (\XX) \wgrd (\XX)^n ] = \esp [ \fnl (\XX) {\textstyle \prod_{i=1}^n \tilde{\wgrd} (\XX;\YY\!_i)} ]
\end{equation}
with the $\YY\!_i$ all independent and identically distributed. This also extends to analytic functions \cite{Dauchet2012, Farges2014}, as in a Monte-Carlo calculation an infinite sum is nothing other than an integral.

\paragraph{We solve the Boltzmann equation}\label{sec:intro-phys-bkw} by using the statistical independence to handle the non-linearity of the collision term, as shown in the previous Par.~\ref{sec:intro-trick}. The Boltzmann equation is previously turned into a purely integral form, from which the needed samplings are deduced. 
Let us see how it works on a simple example.


We consider a case with no spatial dependencies: the gas is uniform, without border, and under the influence of no external force, so $\ff \equiv \ff (\cc;t)$. The gas has unit density. The problem is the estimation of $\ff$ at any given $(\cc;t \geqslant 0)$, given $\ff$ at $t=0$. Let us consider the collision model of Maxwell molecules in the case of isotropic scattering: the differential cross section $\sedf$ is inversely proportional to the relative speed $\ccrn$, with no angular dependence, so that $\ccrn \,\sedf$ is constant. With the adequate non-dimensionalization of time, the Boltzmann equation reads \cite{Krook1976}
\begin{subequations}
\label{eq:bkw-master}
\begin{equation}
\label{eq:bkw-master-time}
\partial_t \ff (\cc;t) = \sfo - \ff (\cc;t) + \sis (\cc;t)
\end{equation}
with
\begin{equation}
\label{eq:bkw-master-src}
\sis (\cc;t) = \int_\spcc \ud\cce \int_\spcu \dfrac{\ud\uup}{4\pi} \, \ff (\ccp;t) \ff (\ccpe;t).
\end{equation}
\end{subequations}
Here the particle derivative $D$ reduces to $\partial_t$, and the ballistic trajectories are only points $\cc$ in the space of speeds. Using the kernel of $\partial_t + \id$ (which is spanned by the decreasing exponential), Eq.~\eqref{eq:bkw-master-time} can be formally solved to
\begin{multline}
\label{eq:bkw-master-tint}
\forall t \geqslant 0, \ff (\cc;t) = \exp (-t) \ff (\cc;0) + \sfo \\
\int_0^t \ud t' \, \exp [-(t-t')] \sis (\cc;t').
\end{multline}

Equations~\eqref{eq:bkw-master-tint} and \eqref{eq:bkw-master-src} give together a recursive non-linear integral equation for $\ff$. By setting a random variable for each of the integration variables $t'$, $\cce$, and $\uup$, this can be turned into the expression of a statistical expectation \cite{DelaTorre2014}:
\begin{multline}
\label{eq:bkw-master-stat}
\ff (\cc;t) = \Pr (T' \leqslant 0) \dfrac{\exp(-t) \ff(\cc;0)}{\Pr (T' \leqslant 0)} + \sfo \\
\int_0^t \pdf{T'} (t') \ud t' \int_\spcc \pdf{\CCE} (\cce) \ud\cce \int_\spcu \pdf{\UUP} (\uup) \ud\uup \times \sfo \\[1ex]
\dfrac{\exp [-(t-t')] \, \ff (\ccp;t') \, \ff (\ccpe;t')}{4\pi \, \pdf{T'}(t') \,\pdf{\CCE}(\cce) \,\pdf{\UUP}(\uup)},
\end{multline}
where the RV $T'$ has its values in the interval $(-\infty \intersep t]$, or equivalently in the interval $[0 \intersep t]$. This leads readily to Monte-Carlo Algo.~\ref{algo:bkw-simple}, which estimates $\ff (\cc;t)$.

\begin{algorithm}[!ht]
	\SetAlgoVlined
	\SetKwFunction{estimf}{$\Ff_{\algo 1}$}
	\KwFunc{\estimf $(\cc;t)$}
	\BlankLine
	\KwIn{A point $(\cc;t)$ in $\spcc \times \reels^+$}
	\BlankLine
	\KwOut{An estimate of $\ff(\cc;t)$}
	\BlankLine
	\BlankLine
	Sample $T'$ in $(-\infty \intersep t]$: $t'$ is obtained\;
	\If{$t' \leqslant 0$}{
		\Return{$\dfrac{\exp (-t) \, \ff(\cc;0)}{\Pr (T' \leqslant 0)}$}\; 
		\tcp{Ends the recursion}
	}
	\Else{
		Sample $\CCE$ in $\spcc$: $\cce$ is obtained\;
		Sample $\UUP$ in $\spcu$: $\uup$ is obtained\;
		$\ccp \gets \tfrac{1}{2} ( \cc + \cce + \| \cc - \cce \| \,\uup )$\;
		$\ccpe \gets \tfrac{1}{2} ( \cc + \cce - \| \cc - \cce \| \,\uup )$\;
		\tcp{Recursively estimate  $\ff (\ccp;t')$ and $\ff (\ccpe;t')$}
		$\Ff_1 \gets \estimf (\ccp;t')$\;
		$\Ff_2 \gets \estimf  (\ccpe;t')$\;
		\Return{$\dfrac{\exp [-(t-t')] \, \Ff_1 \, \Ff_2}{4\pi \, \pdf{T'}(t') \,\pdf{\CCE}(\cce) \,\pdf{\UUP}(\uup)}$}\;
	}
\parbox{\linewidth}{\caption{Estimation of $\ff(\cc;t)$, under the conditions described in Par.~\ref{sec:intro-phys-bkw} (uniform gas with unit density, Maxwell molecules\dots). $\ff(\cc;0)$ is supposed to be known, and $\Pr (\dotso)$ means ``probability''. This algorithm is recursive.}}
\label{algo:bkw-simple}
\end{algorithm}

Following Eq.~\eqref{eq:espsquare}, $\ff (\ccp;t')$ and $\ff (\ccpe;t')$ in Algo.~\ref{algo:bkw-simple} are estimated independently, by applying the same Algo.~\ref{algo:bkw-simple} further backwards in time, up to the initial condition. At each time sampling, the initial time can be reached, in which case $\ff$ is estimated by the initial condition and the recursion stops. If the samplings are settled correctly, the algorithm will surely terminate.

The recursive character of Algo.~\ref{algo:bkw-simple} is a feature common to all Monte-Carlo algorithms solving multiple scattering problems of transport physics. Here, however, due to the non-linearity, each recursive stage is likely to call itself more than once (twice, in this precise case). So, to be exact, the estimation process no longer consists in following a particle path, but rather in following a particle branching tree. In the following article, we will use the term ``particle tree'' as often as ``estimation tree'', because at each particle position in the phase space-time the distribution function would be estimated.
We will also call the probe point where $\ff (\cc;t)$ is finally estimated the ``root'' of an estimation tree, each ballistic trajectory associated with a $T'$ sampling is an ``edge'' of this tree, and the nodes of a tree where it meets the initial condition and the recursion stops are the ``leaves'' of the tree [depicted in Fig.~\ref{fig:differents-suivis}(b) by red crosses].

\paragraph{Algorithm~\ref{algo:bkw-simple},} as it is presented, estimates $\ff (\cc;t)$ only once. Estimating the statistical noise (i.e., the variance) associated with the estimation of $\ff (\cc;t)$, and thus assessing the reliability of such an estimation, requires $N > 1$ realizations of Algo.~\ref{algo:bkw-simple}. These different realizations are independent from each other, as is usual in the MCM applied to linear transport physics. This makes a difference with the numerical methods of the DSMC family, in which the particles are a swarm which is simulated all at once and collide each other \cite{Bird1994, Bird1998, Oran1998, Murrone2011, Khisamutdinov2004, Stefanov2011, Homolle2007, Degond2010}.

As it is described, Algo.~\ref{algo:bkw-simple} is intended to evaluate the distribution function at a prescribed probe point of the phase space (what is called a ``probe calculation''). It does not output and does not compute the distribution function in the whole phase space before the probe date, as is done in methods with full discretization of the phase space (such as the Lattice Boltzmann methods \cite{Piaud2014, Ambrus2016, Sofonea2003, He1997a, He1997b, Chen1998}, Unified Gas Kinetic Scheme \cite{Chen2012a, Huang2012, Guo2013}, Fast Kinetic Scheme \cite{Dimarco2013, Dimarco2013b, Dimarco2015}, or Discrete Velocity Methods \cite{Sharipov1993, Palczewski1997, Palczewski1998, Mieussens2000}), or in statistical methods of the DSMC family \cite{Bird1994, Bird1998, Oran1998, Khisamutdinov2004, Homolle2007, Degond2010, Stefanov2011, Murrone2011}. Instead of this, it will estimate the distribution function at a few locations in this phase space-time before the probe, where collisions need to be sampled. This ability to perform probe estimation has interesting consequences, which we will illustrate below.

\section{Application: the BKW mode}
\label{sec:bkw}

\paragraph{Equation~\eqref{eq:bkw-master} admits an explicit solution,} called the Bobylev-Krook-Wu mode. It was discovered by these authors in 1976 \cite{Bobylev1976, Krook1976} and by Krupp in 1967 \cite{Krupp1967}. 

\begin{figure}[htb]
\centering
\includegraphics{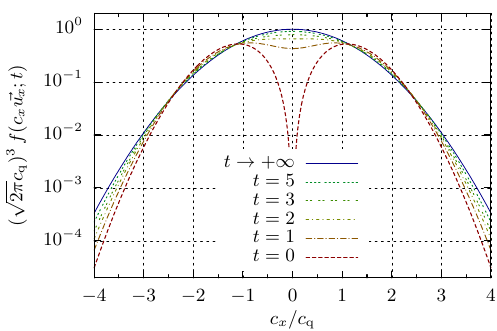}
\caption{The distribution function $\ff$ along an axis of the velocity space, according to the BKW mode described in Eq.~\eqref{eq:expr-mode-bkw}, at several times $t$. The BKW mode is isotropic, i.e., $\ff \equiv \ff (\| \cc \|;t)$: $\ux$ is any unit vector.
}
\label{fig:bkw-mode}
\end{figure}

If $\spcc$ is three-dimensional, this solution states that (see \cite{Ernst1979} for extension of the BKW mode to any dimensionality)
\begin{subequations}
\label{eq:expr-mode-bkw}
\begin{equation}
\ff (\cc;t) = \dfrac{\exp [-\ccd / (2\bkwk\cqm^2)]}{2(\sqrt{2\pi\bkwk}\cqm)^3} \biggl( \dfrac{5\bkwk - 3}{\bkwk} + \dfrac{1-\bkwk}{\bkwk^2} \times \dfrac{\ccd}{\cqm^2} \biggr),
\end{equation}
with
\begin{equation}
\bkwk = 1 - \tfrac{2}{5} \exp (-t/6)
\end{equation}
\end{subequations}
and $\cqm$ the Root-Mean-Square velocity on each axis (RMS velocity). With the $\frac{2}{5}$ factor in the expression of $\bkwk$, $\ff$ is finite and positive for $t \geqslant 0$ (texts in the references adopt different conventions). In short, the BKW mode describes the relaxation from a particular initial disequilibrium toward an equilibrium. This equilibrium, corresponding to $t \to +\infty$ and $\bkwk = 1$, is the well-known Maxwell distribution
\begin{equation}
\label{eq:bkw-equilibrium}
\ff_\eq (\cc) = \dfrac{\exp [ - \ccd / (2\cqm\sfo^2) ]}{(\sqrt{2\pi}\cqm)^3}.
\end{equation}
The $\frac{2}{5}$ factor in the expression of $\bkwk$ sets a initial disequilibrium maximal accounting for the positivity constraint on $\ff$. Then
\begin{equation}
\label{eq:bkw-max-disequilibrium}
\ff (\cc;0) = \dfrac{5\,\ccd}{9\,\cqm^2} \, \biggl( \sqrt{\dfrac{6\pi}{5}} \, \cqm \biggr)^{\!\!-3} \exp \biggl[ - \ccd \bigg/ \biggl( \dfrac{6}{5} \, \cqm^2 \biggr) \biggr].
\end{equation}
Figure~\ref{fig:bkw-mode} gives a graphical overview of the BKW mode.

\paragraph{}\label{sec:algo-2-sampling-and-properties} Since the physical case in which the BKW mode stands is the one used in Sec.~\ref{sec:principle}, Algo.~\ref{algo:bkw-simple} can be applied to it. Before setting up a real test, we only have to specify the sampling laws for $t'$, $\cce$, and $\uup$.

We propose the following choice:
\begin{equation}
\label{eq:probas-simples-bkw}
\left\{\begin{aligned}
\pdf{T'} &: \left\{\begin{array}{@{}l}
(-\infty \intersep t] \to \reels^{+*} \\
t' \mapsto \exp [-(t-t')] \end{array}\right.\\
\pdf{\CCE} &: \left\{\begin{array}{@{}l}
\spcc \to \reels^{+*} \\
\cce \mapsto \ff_\eq (\cce) \end{array}\right.\\
\pdf{\UUP} &: \left\{\begin{array}{@{}l}
\spcu \to \reels^{+*} \\
\uup \mapsto (4\pi)^{-1} \end{array}\right.
\end{aligned}\right.,
\end{equation}
i.e., $T'$ follows a unit-scale exponential law for $T' \leqslant t$, $\CCE$ has the final equilibrium distribution (i.e., a Maxwell distribution with zero peculiar velocity and $\cqm$ RMS speed), while $\UUP$ is isotropically distributed.
Algorithm~\ref{algo:bkw-simple} with these sampling laws will be denoted hereafter Algo.~\ref{algo:bkw-simple}-bkw.

\begin{figure*}[!ht]
\parbox{\Graphwidth}{\includegraphics{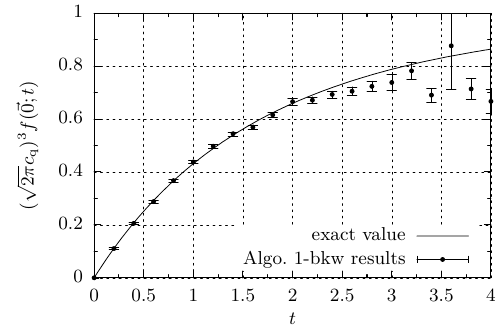}
\hfill
\includegraphics{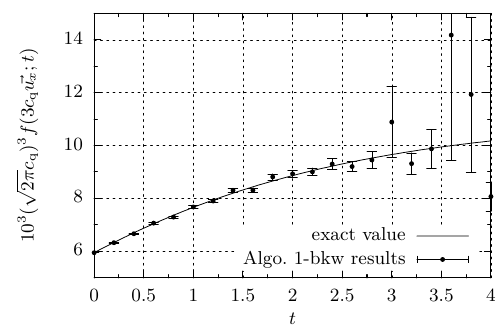}}
\\ \vspace{\baselineskip}
\parbox{\Graphwidth}{\includegraphics{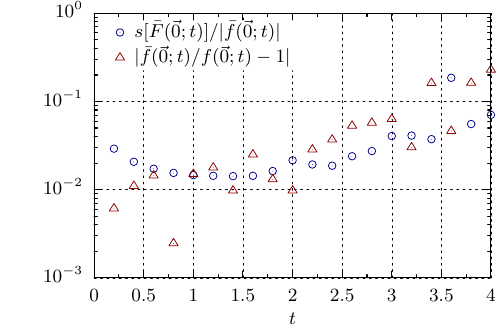}
\hfill
\includegraphics{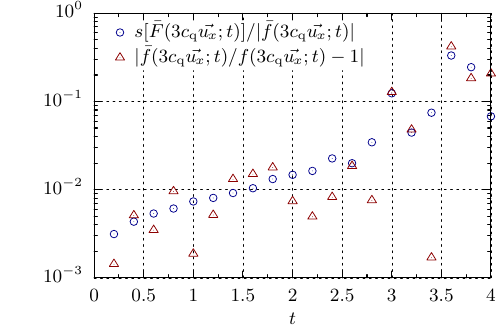}}
\\
\caption{Results obtained by Algo.~\ref{algo:bkw-simple}-bkw, applied to the BKW mode described in Eq.~\eqref{eq:expr-mode-bkw}. We follow the distribution function $\ff$ along the time $t$, at two probe points of the velocity space: the center of the velocity space in graphs in the left column, and a point located $3\times$ the RMS velocity away from that center in graphs in the right column. The graphs at the top give a graphical comparison between the results (given with confidence intervals of 1 standard deviation) and the expected values. The graphs at the bottom show, with the same results, the relative standard deviation (blue circles, given by the calculations) besides the error actually made (red triangles). Each displayed point was obtained through running \num{e4} realizations of Algo.~\ref{algo:bkw-simple}-bkw.}
\label{fig:bkw-res-pos-fixe}
\end{figure*}

\begin{figure*}[!ht]
\parbox{\Graphwidth}{\includegraphics{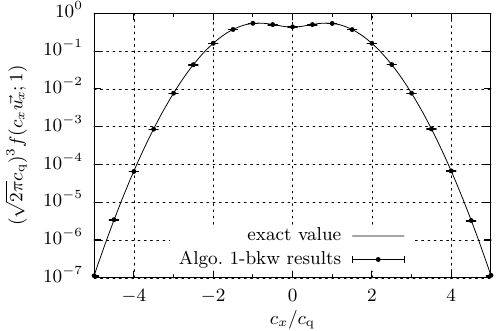}
\hfill
\includegraphics{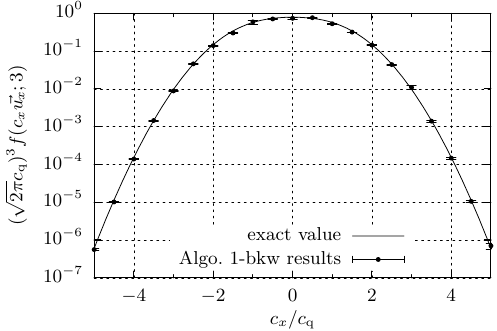}}
\\ \vspace{\baselineskip}
\parbox{\Graphwidth}{\includegraphics{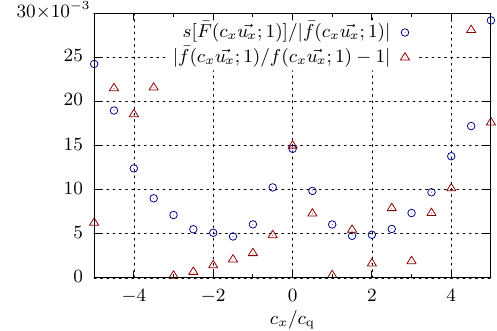}
\hfill
\includegraphics{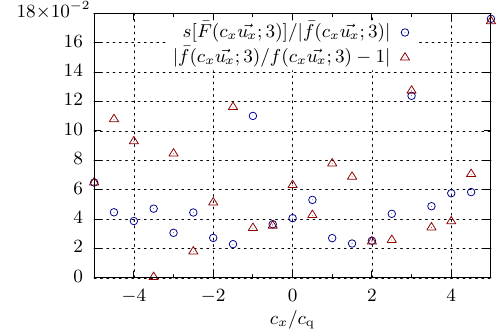}}
\\
\caption{Results obtained by Algo.~\ref{algo:bkw-simple}-bkw, applied to the BKW mode described in Eq.~\eqref{eq:expr-mode-bkw}. Probe points are spread on an $(Ox)$ axis of the velocity space [$\ux$ is the unit vector of $(Ox)$], and two probe dates are considered: 1 mean collision time after the initial condition in graphs in the left column, and 3 mean collision times in graphs in the right column. The graphs at the top give a graphical comparison between the results (given with confidence intervals of 1 standard deviation) and the expected values. The graphs at the bottom show, with the same results, the relative standard deviation (blue circles, given by the calculations) besides the error actually made (red triangles). Each displayed point was obtained through running \num{e4} realizations of Algo.~\ref{algo:bkw-simple}-bkw.}
\label{fig:bkw-res-t-fixe}
\end{figure*}

This choice, although guided by a desire for simplicity, brings two interesting qualities to Algo.~\ref{algo:bkw-simple}-bkw:
\begin{enumerate}
\item The expectation of the recursion, defined as the number of $T'$ samplings per final estimation of $\ff(\cc;t)$ (i.e., the number of ballistic trajectories in a particle tree), is finite. This implies that Algo.~\ref{algo:bkw-simple}-bkw surely terminates.
\item If at $t=0$, $\ff$ was the equilibrium distribution
\begin{equation}
\ff(\cc;0) = \ff_\indic{eq} (\cc)\texte{, i.e., } \bkwk = 1,
\end{equation}
then Algo.~\ref{algo:bkw-simple}-bkw would have a null variance, i.e., each of its executions would output exactly the expected result. 
\end{enumerate}

One may be confident in the above statements for the following reasons:
\begin{proof}[1\textsuperscript{st} point]
Let us consider the expectation of the recursion of Algo.~\ref{algo:bkw-simple}-bkw starting at $(\cc;t)$, defined as the expectation of the number of times $T'$ random variables will be sampled. This defines a function $\rec: \spcc \times \reels^+ \to [1 \intersep +\infty]$.

$\rec(\cc;t)$ follows an integral expression similar to Eq.~\eqref{eq:bkw-master-stat}. To obtain such an expression, one has to take back Eq.~\eqref{eq:bkw-master-stat} and to replace the estimator of $\ff$ by a mean count of $T'$ samplings. This mean count always counts 1 as Algo.~\ref{algo:bkw-simple} starts by a $T'$ sampling, plus the mean counts of $T'$ samplings in the recursive calls of Algo.~\ref{algo:bkw-simple} if these are needed. This means that
\begin{multline}
\label{eq:bkw-rec-int-pdfs}
\rec (\cc;t) = 1 + \Pr (T' \leqslant 0) \times 0 + \sfo \\
\int_0^t \pdf{T'}(t') \ud t' \int_\spcc \pdf{\CCE}(\cce) \ud\cce \int_\spcu \pdf{\UUP}(\uup) \ud\uup \times \sfo \\
[ \rec (\ccp;t') + \rec (\ccpe;t') ].
\end{multline}
Then, accounting for the sampling choices listed in Eqs.~\eqref{eq:probas-simples-bkw}, one gets
\begin{multline}
\label{eq:bkw-rec-int-details}
\rec (\cc;t) = 1 + \sfo \\
\int_0^t \exp [-(t-t')] \ud t' \int_\spcc \dfrac{\exp [-\cce\sfo^2/(2\,\cqm^2)] \ud\cce}{\bigl(\sqrt{2\pi}\cqm\bigr)^3} \int_\spcu \dfrac{\ud\uup}{4\pi} \times \sfo \\
[ \rec (\ccp;t') + \rec (\ccpe;t') ].
\end{multline}

In order to solve this integral equation explicitly, we take the derivative of both sides with respect to $t$. One obtains thus
\begin{equation}
\label{eq:bkw-rec-diff}
\left\{\begin{array}{@{\ }l@{}}
\forall t \geqslant 0, \left\{\begin{multlined}
\partial_t \rec (\cc;t) = \sfo - \rec (\cc;t) + 1 + \sfo \\
\displaystyle \int_\spcc \dfrac{\exp [-\cce\sfo^2/(2\,\cqm^2)] \ud\cce}{(\sqrt{2\pi}\cqm)^3} \int_\spcu \dfrac{\ud\uup}{4\pi} \times \sfo \\
[ \rec (\ccp;t) + \rec (\ccpe;t) ],
\end{multlined}\right.\\
\\[0ex]
\rec (\cc;t=0) = 1.
\end{array}\right.
\end{equation}
$\rec$ is uniform at $t=0$, and its time derivative preserves this uniformity: so here $\rec (\cc;t) \equiv \rec (t)$. Finally the system is easily solved:
\begin{equation}
\label{eq:bkw-rec-result}
\rec (\cc;t) = 2 \exp (t) - 1,
\end{equation}
which is finite for any $(\cc;t) \in \spcc \times \reels^+$.
\end{proof}
\begin{proof}[2\textsuperscript{nd} point]
Let $\FF (\cc;t)$ be the estimator of $\ff (\cc;t)$ obtained through Algo.~\ref{algo:bkw-simple}-bkw. The initial distribution $\ff (\cc; 0)$ is assumed to be the equilibrium distribution described in Eq.~\eqref{eq:bkw-equilibrium}.

Now consider the following induction hypothesis, to be applied anywhere in an estimation tree: $\FF(\cc;t) = \ff_\eq (\cc) = (\sqrt{2\pi}\cqm)^{-3} \exp [- \ccd / (2\, \cqm^2)]$, i.e., $\ff$ is always estimated exactly as the equilibrium distribution, with zero variance.

Starting Algo.~\ref{algo:bkw-simple}-bkw there are two possibilities: $T' \leqslant 0$  or $T' > 0$.

If $T' \leqslant 0$, a leaf of the estimation tree is reached. Here takes place the initialization of our proof by induction. Indeed,
\begin{align}
\label{eq:bkw-zerovar-init}
\FF (\cc;t) &= \ff(\cc;0) \notag \\
 &= \ff_\eq (\cc),
\end{align}
the induction hypothesis is true.

If $T' > 0$, a non-leaf node of the tree is reached. Then
\begin{align*}
\FF (\cc;t) &= \dfrac{(\sqrt{2\pi}\cqm)\sfo^3 \, \FF_1 \FF_2}{\exp [-\CCE\sfo^2/(2\,\cqm^2)]} \\[1ex]
 &= \dfrac{(\sqrt{2\pi}\cqm)\sfo^3 \, \FF (\CCP;T') \, \FF (\CCPE;T')}{\exp [-\CCE\sfo^2/(2\,\cqm^2)]},
\end{align*}
where $\FF_1$, $\FF_2$, $\CCP$, and $\CCPE$ are the RVs corresponding to the values $\Ff_1$, $\Ff_2$, $\ccp$, and $\ccpe$ in Algo.~\ref{algo:bkw-simple}-bkw. 
If the induction hypothesis holds for $\FF (\CCP;T')$ and $\FF (\CCPE;T')$, then
\begin{align}
\label{eq:bkw-zerovar-recur}
\FF (\cc;t) &= \dfrac{ \exp [-\CCP\sfo^2/(2\,\cqm^2)] \, \exp [-\CCPE\sfo^2/(2\,\cqm^2)]}{(\sqrt{2\pi}\cqm)^3 \,\exp [-\CCE\sfo^2/(2\,\cqm^2)]} \notag \\
 &= \dfrac{\exp [-\ccd/(2\,\cqm^2)]}{(\sqrt{2\pi}\cqm)^3},
\end{align}
because $\CCP\sfo^2 + \CCPE\sfo^2 = \ccd + \CCE\sfo^2$. The induction hypothesis propagates.

Because the estimation tree is finite (cf.\@ the previous point), the induction hypothesis applies from the leaves of the tree to any node of the tree, root included.
\end{proof}

\paragraph{Numerical experiments}\label{sec:bkw-result-comments} were carried out in which Algo.~\ref{algo:bkw-simple}-bkw was operated to calculate $\ff$ at several points of the phase space-time. The physical situation considered is the BKW mode expressed in Eq.~\eqref{eq:expr-mode-bkw}.
For each point where $\ff$ was calculated, \num{e4} realizations of Algo.~\ref{algo:bkw-simple}-bkw were used. The results $\bar\ff$ are displayed in Figs.~\ref{fig:bkw-res-pos-fixe} and \ref{fig:bkw-res-t-fixe}.

A first feature of the performance of Algo.~\ref{algo:bkw-simple}-bkw is that the estimate's error increases with the time elapsed since the initial condition. This can be explained in the following ways:
\begin{itemize}
\item When the elapsed time $t$ increases, on average the estimation trees get more nodes and leaves [$\rec (t)$ counts, indeed, the expected number of lines in the estimation trees.]. Because the relation between the result and the equilibrium value is exactly the product of these relations at every leaf of the estimation tree (this can be understood through the justification of the 2\textsuperscript{nd} property of Algo.~\ref{algo:bkw-simple}-bkw, listed in the previous Par.~\ref{sec:algo-2-sampling-and-properties}), more leaves mean stronger variance in the final result.

\item Consider two independent RVs $X$ and $Y$ with real values, then
\begin{subequations}
\begin{equation}
\esp (XY) = \esp (X) \esp (Y)
\end{equation}
and
\begin{equation}
\esp (X^2 Y^2) = \esp (X^2) \esp (Y^2),
\end{equation}
so
\begin{equation}
\dfrac{\var (XY)}{\esp (XY)^2} + 1 = \biggl( \dfrac{\var (X)}{\esp (X)^2} + 1 \biggr) \biggl( \dfrac{\var (X)}{\esp (X)^2} + 1 \biggr).
\end{equation}
\end{subequations}
This implies that if two independent RVs have low (resp. high) relative variances, the relative variance of their product is the sum (resp. product) of their relative variances. The result of Algo.~\ref{algo:bkw-simple}-bkw is a product of independent realizations of itself during a branching process, each of these intermediate estimations having a variance. As the number of intermediate estimations increases exponentially with simulated time $t$ [as demonstrates the evolution of $\rec (t)$], the variance of the final result increases with $t$.

\item The modeling process we applied to the mean recursion $\rec(\cc;t)$ through Eqs.~\eqref{eq:bkw-rec-int-pdfs} to \eqref{eq:bkw-rec-diff} can be applied to the estimator variance (this will be the subject of future work), which does indeed show that this variance increases with time $t$. The very reason is the one exposed in the previous point: the relative variance of the intermediate estimations adds and multiplies.
\end{itemize}

A more interesting behavior of Algo.~\ref{algo:bkw-simple}-bkw is that the variance of its result depends very little on the position in the phase space. In particular, it is nearly independent of the rarefaction: Fig.~\ref{fig:bkw-res-t-fixe} shows that at location in the velocity space where $\ff$ is \num{e6} smaller than at the center of $\spcc$, the relative error of the obtained estimations increases only by a factor of 5. This makes an important difference with the methods of the DSMC family, with which it is impossible to estimate the density $\ff$ in such empty parts of the phase space (the relative variance of the estimations gets too strong). Typically, to estimate the fraction $\Frac$ of molecules inside a part of the phase space, the DSMC methods need to track much more than $1 / \Frac$ molecules to obtain an estimation with a reasonable variance; with our method, the relative variance of such an estimation seems nearly insensitive to the scarcity of molecules in the probed region.

In order to verify this last feature in more depth, we computed high-energy fractions of the gas. The quantity we have calculated is $\Frac(\| \cc \| \geqslant c_0; t)$, the fraction of total number of molecules that have a speed exceeding a given value $c_0$ at time $t$; it is the complementary cumulative distribution function (or tail distribution) of $\| \cc \|$. 
Since it is an integral of $\ff$ on $\spcc$, it can be computed using Monte-Carlo Algo.~\ref{algo:HS-integrator}, encapsulating Algo.~\ref{algo:bkw-simple}-bkw with an additional sampling of a final velocity $\CC_\final = C_\final \,\UU_\final$, with
\begin{equation}
\label{eq:HS-sampling}
\left\{\begin{array}{@{\ }l@{\ \ }l@{}}
\pdf{C_\final} (c_\final) = \dfrac{c_0 \Bigl[ \Bigl(\dfrac{c_0}{\cqm}\Bigr)^{\!2} - 2\Bigr]}{\Bigl\{ \Bigl[ \Bigl(\dfrac{c_0}{\cqm}\Bigr)^{\!2} - 2 \Bigr] (c_\final - c_0) + c_0 \Bigr\}^{\!2}} & \texte{if } c_0 > 2\cqm, \\ \\[-1ex]
\pdf{C_\final} (c_\final) = \dfrac{2 \cqm}{(2 \cqm + c_\final - c_0)^2} & \texte{if } c_0 \leqslant 2\cqm, \\ \\[-1ex]
\UU_\final \texte{ of isotropic law}. &
\end{array}\right.
\end{equation}

\begin{algorithm}[!ht]
	\SetAlgoVlined
	\KwIn{A speed $c_0$ and a time $t$}
	\BlankLine
	\KwOut{An estimate of $\Frac(\| \cc \| \geqslant c_0; t)$, the fraction of particles the speed of which exceeds $c_0$ at time $t$}
	\BlankLine
	Sample $C_\final$: $c_\final$ is obtained\tcp*[r]{$c_\final \geqslant c_0$}
	Sample $\UU_\final$: $\uu_\final$ is obtained\;
	Estimate $\ff (c_\final \mspace{1mu} \uu_\final; t)$ using Algo.~\ref{algo:bkw-simple}-bkw: $\Ff$ is obtained\;
	\Return{$\dfrac{c_\final\sfo^2 \, \Ff}{\pdf{C_\final}(c_\final) \, \pdf{\UU_\final}(\uu_\final)}$}\;
\parbox{\linewidth}{\caption{Estimation of the tail distribution of kinetic energies of molecules, valid with the physical conditions described in Secs.~\ref{sec:principle} and \ref{sec:bkw} (no space dependency, Maxwell molecules, etc.)}}
\label{algo:HS-integrator}
\end{algorithm}

\begin{figure*}[th]
\parbox{\Graphwidth}{\includegraphics{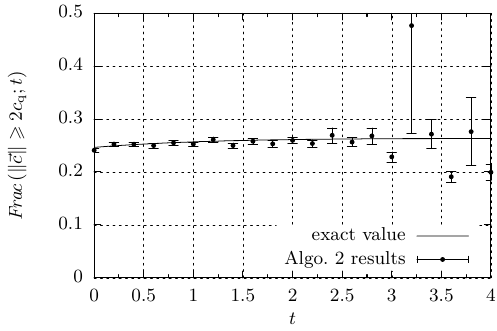}
\hfill
\includegraphics{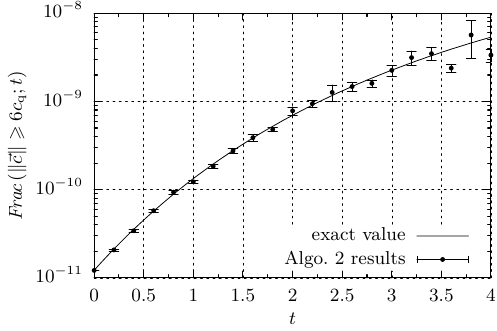}}
\\ \vspace{\baselineskip}
\parbox{\Graphwidth}{\includegraphics{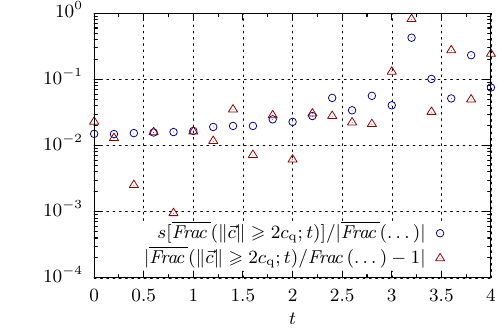}
\hfill
\includegraphics{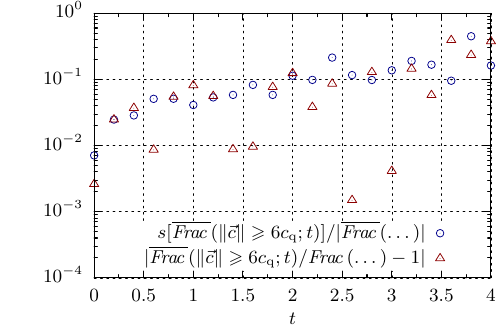}}
\\
\caption{Results obtained by Algo.~\ref{algo:HS-integrator} combined with the sampling law listed in Eq.~\eqref{eq:HS-sampling}, and applied to the BKW mode described in Eq.~\eqref{eq:expr-mode-bkw}. Two constant speeds are considered: along the time $t$, we follow how many particles have a speed above these thresholds. The threshold is $2\times$ the RMS velocity in graphs in the left column, and $6\times$ the RMS velocity in graphs in the right column. The graphs at the top give a graphical comparison between the results (given with confidence intervals of 1 standard deviation) and the expected values. The graphs at the bottom show, with the same results, the relative standard deviation (blue circles, given by the calculations) besides the error actually made (red triangles). Each displayed point was obtained through running \num{e4} realizations of Algo.~\ref{algo:HS-integrator}.}
\label{fig:bkw-res-cm}
\end{figure*}

\begin{figure*}[th]
\parbox{\Graphwidth}{\includegraphics{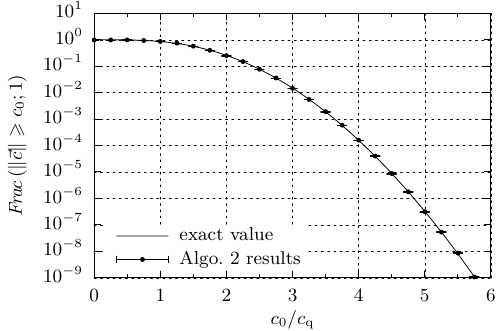}
\hfill
\includegraphics{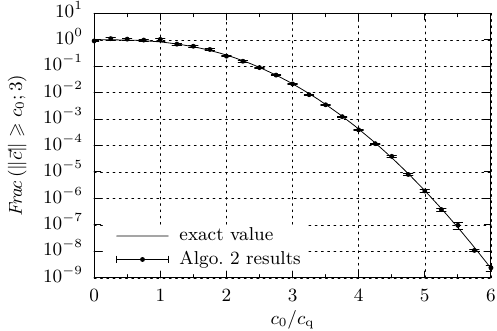}}
\\ \vspace{\baselineskip}
\parbox{\Graphwidth}{\includegraphics{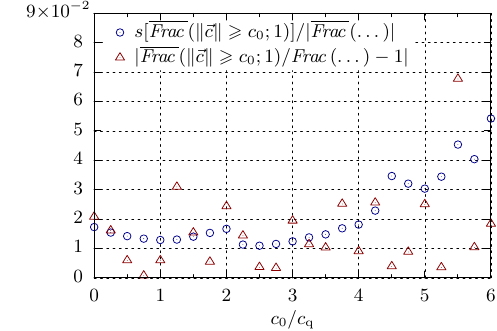}
\hfill
\includegraphics{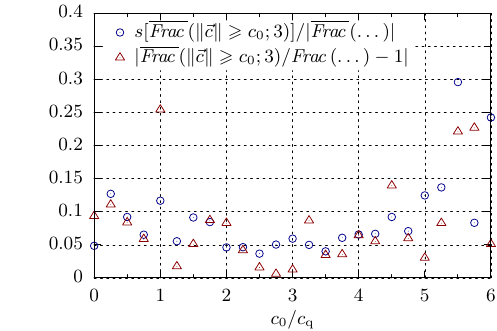}}
\\
\caption{Results obtained by Algo.~\ref{algo:HS-integrator} combined with the sampling laws listed in Eq.~\eqref{eq:HS-sampling}, and applied to the BKW mode described in Eq.~\eqref{eq:expr-mode-bkw}. We calculate the distribution of particles along the speed $\|\cc\|$, i.e., how many particles have a speed above any given threshold $c_0$; two dates are considered: 1 mean collision time after the initial condition in graphs in the left column, 3 mean collision times in graphs in the right column. The graphs at the top give a graphical comparison between the results (given with confidence intervals of 1 standard deviation) and the expected values. The graphs at the bottom show, with the same results, the relative standard deviation (blue circles, given by the calculations) besides the error actually made (red triangles). Each displayed point was obtained through running \num{e4} realizations of Algo.~\ref{algo:HS-integrator}.}
\label{fig:bkw-res-Qft}
\end{figure*}

The results
of Algo.~\ref{algo:HS-integrator}, with sampling laws listed in Eqs.~\eqref{eq:HS-sampling}, are displayed in Figs.~\ref{fig:bkw-res-cm} and \ref{fig:bkw-res-Qft}. There seems to be no problem in quantifying tiny parts of the gas, especially those located at high kinetic energy, even fractions as small as one billionth of the total.

\section{Toy model for NCA}
\label{sec:toy}

\paragraph{In this section we address} how to deal with the non-linearity of the extinction term in the Boltzmann equation. In Secs.~\ref{sec:principle} and \ref{sec:bkw} the collision frequency was constant, due to the combination of the uniform density and the Maxwell collision model. In a more general case, the collision frequency varies depending on $\ff$, and this introduces an additional non-linearity in Eq.~\eqref{eq:bkw-master-tint} through the exponential term. We show here how to handle this using Null Collision Algorithms (NCAs).

To introduce it simply, we isolate the extinction of the Boltzmann dynamics, with no source term nor phase space. This amounts to solving the quadratic ordinary differential equation
\begin{equation}
\label{eq:toy-system}
\left\{ \begin{aligned}
\ft' (t) &= - \alpha\, \ft (t)^2 \\
\ft (0) &= \ft_0
\end{aligned} \right.,
\end{equation}
where $\ff$ is the unknown function, and $\alpha$ and $\ft_0$ are arbitrary positive constants. It admits on $t \in \reels^+$ the only solution
\begin{equation}
\label{eq:toy-solution}
\ft (t) = \dfrac{\ft_0}{\alpha\, \ft_0\, t + 1}.
\end{equation}

\paragraph{A way to convert} Eq.~\eqref{eq:toy-system} into an integral form convertible into a Monte-Carlo algorithm, would be to use a linear solution of Eq.~\eqref{eq:toy-system}, as we did to convert Eq.~\eqref{eq:bkw-master-time} into Eq.~\eqref{eq:bkw-master-tint}. Indeed, here
\begin{equation}
\ft' (t) = - [ \alpha\, \ft (t) ] \times \ft (t),
\end{equation}
and so
\begin{equation}
\label{eq:toy-beer-unknown}
\forall t \geqslant 0, \ft (t) = \exp \biggl( - \alpha \int_0^t \ud t' \,\ft (t') \biggr) \ft_0.
\end{equation}
Combined with the use of independent and identically distributed RVs applied to a Taylor series of the exponential function, as proposed in Par.~\ref{sec:intro-trick} and detailed in \cite{Dauchet2012}, Eq.~\eqref{eq:toy-beer-unknown} leads to a recursive integral writing of $\ft(t)$, which can be turned into a Monte-Carlo algorithm. However, we have not used this directly, but have rather relied on a NCA. NCAs are formally equivalent to the use of a Taylor series of the exponential extinction law; this has been explained, for example, in \cite{Longo2002, Galtier2015b, Georgiev2019}. But the way the NCAs are built relates to a physical interpretation, and their convergence is very simple to obtain, which is why we have chosen this approach.

The principle of the Null Collision technique is to set the extinction frequency to an arbitrary value $\kc$. This is compensated by adding a source term of particles having encountered null collisions (collisions without effect). Equation~\eqref{eq:toy-system} is thus transformed into
\begin{equation}
\label{eq:toy-null-injection}
\ft' (t) = - \kc \ft(t) + [\kc \ft (t) - \alpha \ft (t)^2].
\end{equation}
Considering $-\kc \ft(t)$ as an extinction and $[\kc \ft (t) - \alpha \ft (t)^2]$ as a source term, this leads to the integral expression
\begin{multline}
\label{eq:toy-fredholm-acn}
\forall t \geqslant 0, \ft (t) = \exp (- \kc t) \ft_0 + \sfo \\
\int_0^t \ud t' \, \exp [- \kc (t - t')] [ \kc \ft (t') - \alpha\,\ft(t')^2 ].
\end{multline}
A Monte-Carlo algorithm derived from this last expression would be different from another derived from Eq.~\eqref{eq:toy-beer-unknown} and a Taylor series, even if the underlying transport problems are equivalent.
 
In early implementations of NCAs \cite{Skullerud1968, Andreucci1985, Rees1968}, $\kc$ had to exceed the real extinction frequency everywhere for the algorithms to work. Hence we call $\kc$ the ``raised collision frequency'' hereafter. It has been demonstrated recently that this constraint is generally not mandatory \cite{Galtier2013}, and that $\kc$ only needs to be positive. Recently, it has also been shown that the exact knowledge of the extinction is not mandatory, and that it is enough to have an unbiased estimator of the extinction frequency \cite{Galtier2015b}. Choosing $\kc$ as an upper bound of the real extinction frequency (or of its estimator) nevertheless improves the convergence in most cases \cite{Longo2004, Galtier2015b}.

\paragraph{By setting a probability density}\label{sec:toy-previsions} for the integration variable $t'$ in Eq.~\eqref{eq:toy-fredholm-acn} at every $t \geqslant 0$,  a recursive Monte-Carlo algorithm for evaluating $\ft (t)$ is obtained. We choose this density as the exponential term in Eq.~\eqref{eq:toy-fredholm-acn},

\begin{equation}
\pdf{T'} : \left\{ \begin{array}{@{\ }l}
(-\infty \intersep t] \to \reels^{+*} \\
t' \mapsto \kc \exp [- \kc (t - t')]
\end{array} \right.,
\end{equation}
for the sake of simplicity, and this choice will appear reasonable in this section. The result is Algo.~\ref{algo:toy-simple}.
\begin{algorithm}[!ht]
	\SetAlgoVlined
	\SetKwFunction{estimf}{$\Ff_{\algo 3}$}
	\KwFunc{\estimf $(t)$}
	\BlankLine
	\KwIn{A time $t \geqslant 0$}
	\BlankLine
	\KwOut{An estimate of $\ft (t)$}
	\BlankLine
	\BlankLine
	Sample $T'$ from an exponential law for $T' \leqslant t$, with rate $\kc$: $t'$ is obtained\;
	\BlankLine
	\If{$t' \leqslant 0$}{
		\Return{$\ft_0$}\;
		\tcp{Ends the recursion}
	}
	\Else{
		\tcp {Recursively estimate $\ft (t')$ twice}
		$\Ft_1 \gets \estimf(t')$\;
		$\Ft_2 \gets \estimf (t')$\;
		\Return{$\biggl( 1 - \dfrac{\alpha \, \Ft_1}{\kc} \biggr) \Ft_2$}\;
	}
\parbox{\linewidth}{\caption{Estimation of $\ft (t)$, as described in Eq.~\eqref{eq:toy-system}. This algorithm is recursive.}}
\label{algo:toy-simple}
\end{algorithm}

In the same way as shown in the previous section, it can be proven that the expected recursion of Algo.~\ref{algo:toy-simple} is finite. Defining this expected recursion $\rec (t)$ as the expected total number of $T'$ sampling per estimation of $\ft (t)$ by Algo.~\ref{algo:toy-simple} (as in the previous section), we obtain a very similar result:
\begin{equation}
\label{eq:toy-rec-result}
\rec (t) = 2 \exp (\kc t) - 1.
\end{equation}

The same modeling approach can be applied to the second moment of $\FT (t)$ the estimator of $\ft (t)$ built by Algo.~\ref{algo:toy-simple}, in order to obtain its variance. The example given in this section is simple enough to derive an explicit expression of this variance (We will give the details of this derivation in a future work, dedicated to the derivation of estimator variance and algorithmic recursion of algorithms presented in this article, and tips to improve them.). It states that if
\begin{subequations}
\label{eq:toy-var-result}
\begin{equation}
( \alpha\,\ft_0 - \kc ) \,\alpha\,\ft_0\,t < \kc,
\end{equation}
then
\begin{equation}
\var [\FT (t)] = \dfrac{\alpha \, \ft_0 \, t}{\dfrac{\kc}{\alpha\,\ft_0} + ( \kc - \alpha\,\ft_0 ) \,t} \,\ft(t)^2,
\end{equation}
else $\var [\FT (t)]$ is infinite.
\end{subequations}

Interestingly, the behavior of $\FT$ varies qualitatively with the quality of $\kc$ as a global bound of the extinction frequency (of which the maximum is $\alpha \, \ft_0$):
\begin{itemize}
\item If $\kc > \alpha\,\ft_0$, the relative variance of $\FT(t)$, defined as $\var [\FT(t)] \big/ \ft(t)^2$, tends to a finite value when $t \to +\infty$.
$\lim_{t \to + \infty} \var [\FT (t)] = 0$.
$\FT(t)$ is a usable estimator of $\ft(t)$ for every $t$.
\item If $\kc = \alpha\,\ft_0$, the relative variance of $\FT(t)$ grows linearly to infinity with $t$.
$\lim_{t \to + \infty} \var [\FT (t)] = 0$.
$\FT(t)$ is here also a reasonable estimator of $\ft(t)$ regardless of $t$.
\item If $\kc < \alpha\,\ft_0$, the variance of $\FT(t)$ (relative or not) reaches infinity for a finite value of the time $t$, and for all subsequent times. In these conditions, obtaining the convergence of a Monte-Carlo procedure estimating $\ft(t)$ can be very difficult.
\end{itemize}

\begin{figure*}[!htp]
\parbox{\Graphwidth}{\includegraphics{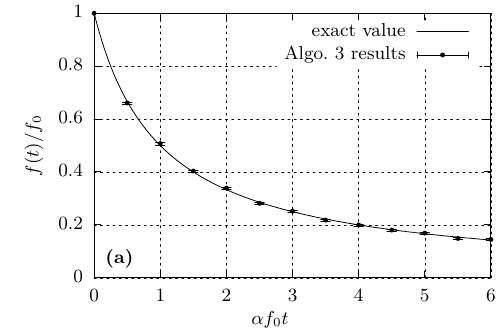}
\hfill
\includegraphics{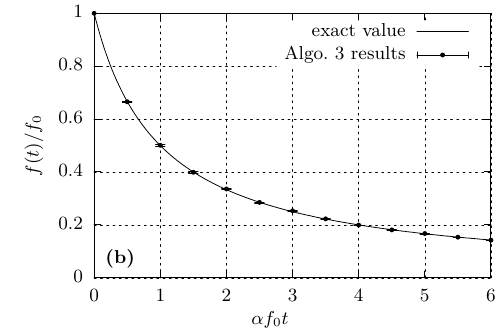}}
\\ \vspace{\baselineskip}
\parbox{\Graphwidth}{\includegraphics{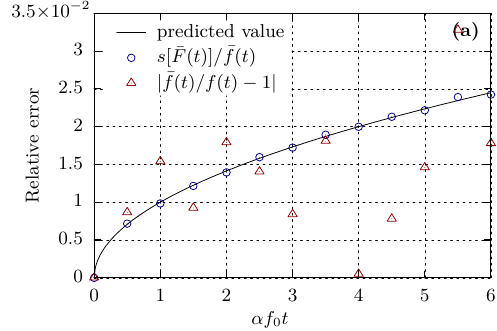}
\hfill
\includegraphics{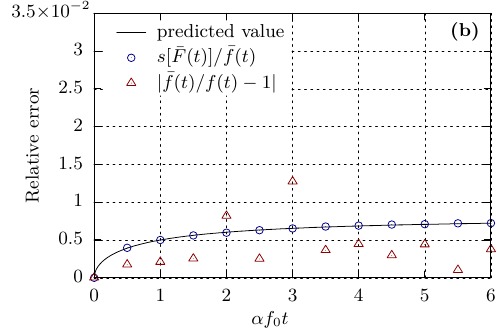}}
\\ \vspace{\baselineskip}
\parbox{\Graphwidth}{\includegraphics{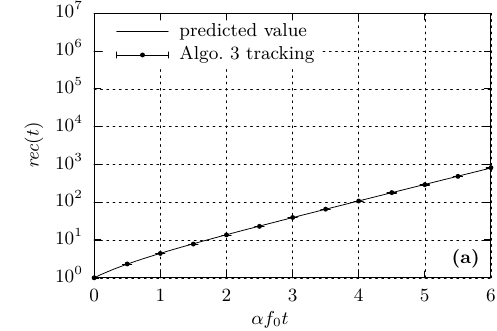}
\hfill
\includegraphics{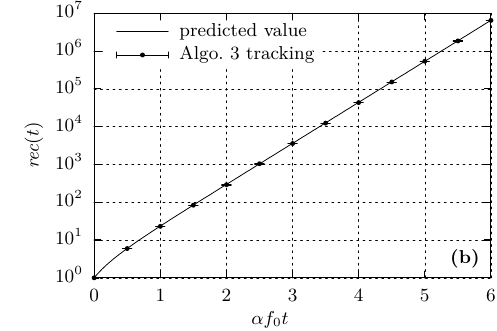}}
\\
\caption{Results obtained by Algo.~\ref{algo:toy-simple}, solving Eq.~\eqref{eq:toy-system}. $\ft(t)$ was calculated at different $t$, using two different values of $\kc$ in columns~(a) ($\kc = 1\,\alpha\,\ft_0$) and (b) ($\kc = \num{2.5}\,\alpha\,\ft_0$).
The four bottom graphs compare the predictions in Eqs.~\eqref{eq:toy-rec-result} and \eqref{eq:toy-var-result} with what Algo.~\ref{algo:toy-simple} has done. The confidence intervals displayed in top and bottom graphs take account of 1 standard deviation. Each point was obtained through running \num{e4} realizations of Algo.~\ref{algo:toy-simple}.}
\label{fig:edo-res}
\end{figure*}

\paragraph{Algorithm~\ref{algo:toy-simple} was actually run,} and some results are given in Fig.~\ref{fig:edo-res}. The statements given in the previous Par.~\ref{sec:toy-previsions} about the recursion and the variance of Algo.~\ref{algo:toy-simple} are confirmed. Noticeably, there is no sharp increase of the estimator variance with time, as is observed in the test on the BKW mode, as much as $\kc$ is set greater than $\alpha\,\ft_0$.

\section{Full example: the harmonic trap}
\label{sec:ht}

\paragraph{In this section the Boltzmann equation is solved}\label{sec:ht-physical situation} using the MCM, in a second academic case where an explicit solution is available. In this case, the gas is no longer uniform, and forms a cloud around the origin which swells and contracts periodically. Compared to the BKW mode, there remains now to account for the ballistic transport and for a variable collision frequency.

The physical case under consideration here is nevertheless very particular, because it belongs to the kernel of the collision operator (i.e., collisions have no influence), although it is not the barometric equilibrium. This possibility was known by Boltzmann himself \cite{Boltzmann1909}, and has been revisited recently in the field of cold atom gas manipulation \cite{Guery-Odelin2014}.

\begin{figure}[!ht]
\includegraphics{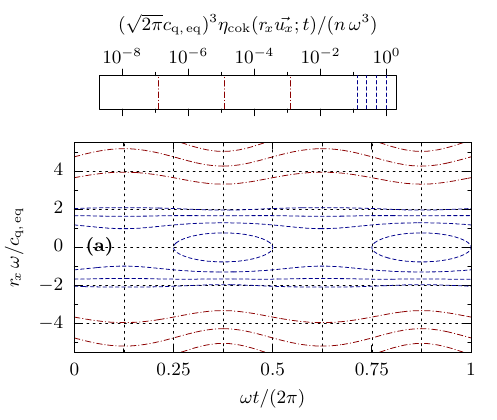}\\
\includegraphics{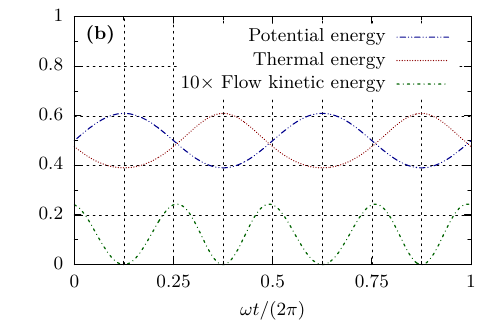}\\
\caption{Presentation of the physical situation of the harmonic trap described in Eq.~\eqref{eq:ht-moments}. The parameters are $\rcqm = \frac{9}{41}$ and $\phaz = 0$, as explained at the end of Par.~\ref{sec:ht-physical situation}. Panel~(a) shows the density of matter along an axis of the ordinary space, as a function of the time (the physical situation has spherical symmetry, so $\ux$ is any unit vector). Panel~(b) shows the distribution of the total mechanical energy of the gas, as a function of the time.}
\label{fig:ht-periodic-breathing}
\end{figure}

Following the details listed in \cite{Guery-Odelin2014}, we built our test case in the collision operator kernel as simply as possible:
\begin{itemize}
\item The molecules are subjected to a stationary and purely elastic force pulling them back to the origin, called a trap. This force yields the acceleration
\begin{equation}
\acc \equiv \acc (\rr) = - \raid^2\,\rr,
\end{equation}
where $\rr$ is the position and $\raid$ a constant angular frequency.
\item The gas has null global kinetic moment around the origin.
\end{itemize}
In these conditions, the equilibrium distributions constitute a set parametrized by two quantities, the total amount of matter $n$ and the thermal RMS speed on each axis $\cqm$. In this case of elastic confinement, the collision operator kernel contains not only the equilibrium distributions; it extends to breathing modes oscillating at twice the trap frequency, with two additional parameters, such as the amplitude of the thermal energy oscillations $\Delta\cqm\!^2$ and a phase at the origin of time $\phaz$. In these breathing modes, the distribution of molecules is always and everywhere Maxwellian, with density, peculiar speed, and thermal energy
\begin{subequations}
\label{eq:ht-moments}
\begin{gather}
\cmol_\cok (\rr;t) = n \, \pdf{\doms{N} \left( \vnul; \tfrac{[1 + \rcqm \sin \phas (t)]\, \cqmeq^2}{\raid^2} \,\matid \right)} (\rr), \\
\Mvv_\cok (\rr;t) = \dfrac{\rcqm \cos \phas (t)}{1 + \rcqm \sin \phas (t)} \, \raid \rr, \\
\cqmcok (t)^2 = \dfrac{(1 - \rcqm^2)\, \cqmeq^2}{1 + \rcqm \sin \phas (t)},
\end{gather}
\end{subequations}
where $\phas (t) = 2 \raid t + \phaz$, $\cqmeq$ is the thermal RMS speed in the equilibrium distribution with the same total mass and the same total mechanical energy, $\rcqm = \frac{\Delta\cqm\!^2}{\cqmeq^2}$ (with the constraint that $0 \leqslant \rcqm < 1$), $\pdf{\doms{N}(\vnul; \doublebar{V})}$ denotes the density of the centered multidimensional normal probability law with covariance matrix $\doublebar{V}$, and $\matid$ is the identity matrix. 
An example of the resulting dynamics is presented in Fig.~\ref{fig:ht-periodic-breathing}, with $\phaz = 0$, and with $\rcqm = \frac{9}{41}$ to comply with the arbitrary constraint that $c_\indic{q,\,min} = \frac{4}{5} c_\indic{q,\,max}$.

\paragraph{Introducing null collisions,}\label{sec:ht-simple-algo} the Boltzmann equation can be written:
\begin{multline}
\label{eq:ht-diff-acn}
\partial_t \ff (\rr;\cc;t) + \cc \cdot \vec{\nabla}_{\!\rr} \,\ff (\rr;\cc;t) - \raid^2 \,\rr \cdot \vec{\nabla}_{\!\cc} \,\ff (\rr;\cc;t) = \\[2ex]
- \frec (\rr;\cc;t) \ff (\rr;\cc;t) + \sfo \\[1ex]
\biggl( \frec (\rr;\cc;t) - \int_\spcc \ud\cce \int_\spcu \ud\uup \ccrn\sedf \ff (\rr;\cce;t) \biggr) \ff (\rr;\cc;t) + \sfo \\
\int_\spcc \ud\cce \int_\spcu \ud\uup \ccrn\sedf \ff (\rr;\ccp;t) \ff (\rr;\ccpe;t),
\end{multline}
where $\sedf \equiv \sedf ( \ccrn; \uup \cdot \uu_\ccr )$, $\ccp$, $\ccpe$, and $\ccr$ are given in Eq.~\eqref{eq:speeds-sphere}, and $\frec$ is the raised collision frequency. Writing the Boltzmann equation under this form requires that the source and loss terms in the collision operator can be split; for other situations see \cite{Degond1991, Villani1998, Gressman2010}. Considering that the distribution is known at $t = 0$, and using Liouville's theorem, Eq.~\eqref{eq:ht-diff-acn} can be turned into a purely integral counterpart,
\begin{subequations}
\label{eq:ht-fred-acn}
\begin{multline}
\ff (\rr;\cc;t) = \exp \biggl( - \int_0^t \ud t' \, \frec [\vecb(t')] \biggr) \ff [\vecb(0)] + \sfo \\
\int_0^t \ud t' \,\exp \biggl( - \int_{t'}^t \ud t'' \, \frec [\vecb(t'')] \biggr) \times \sfo \\
\bm( \{ \frec [\vecb(t')] - \fret [\vecb(t')] \} \ff [\vecb(t')] + \sis [\vecb(t')] \bm),
\end{multline}
where $\vecb (t') = \bm( \rrb(t');\ccb(t');t' \bm)$ describes the ballistic trajectory flowing past $(\rr;\cc)$ at $t$,
\begin{equation}
\left\{\begin{aligned}
\partial_{t'} \rrb (t') &= \ccb (t') \\
\partial_{t'} \ccb (t') &= - \raid^2 \rrb (t')
\end{aligned}\right.
,\ \ \begin{array}{@{}r@{}}
\texte{with the final} \\
\texte{condition}
\end{array}
\ \left\{\begin{aligned}
\rrb (t) &= \rr \\
\ccb (t) &= \cc
\end{aligned}\right.\,,
\end{equation}
and $\fret$ and $\sis$ are the real extinction frequency and source term:
\begin{gather}
\fret (\rr;\cc;t) = \int_\spcc \ud\cce \int_\spcu \ud\uup \, \ccrn \sedf \, \ff (\rr;\cce;t), \\
\sis (\rr;\cc;t) = \int_\spcc \ud\cce \int_\spcu \ud\uup \, \ccrn \sedf \, \ff (\rr;\ccp;t) \ff (\rr;\ccpe;t).\!\!
\end{gather}
\end{subequations}
In the considered harmonic force field, the ballistic trajectories are simple enough to be explicitly computed.

The exponential term expressing the extinction at frequency $\frec$ can be used as a sampling law for $t'$. Then, by choosing the sampling laws for $\cce$ and $\uup$, Eq.~\eqref{eq:ht-fred-acn} can be converted into Monte-Carlo Algo.~\ref{algo:ht-simple}, intended to estimate $\ff(\rr;\cc;t)$ with $(\rr;\cc;t) \in \spcr \times \spcc \times \reels^+$, where $\spcr$ is the space of positions.

\begin{algorithm*}[!tp]
	\SetAlgoVlined
	\SetKwFunction{estimf}{$\Ff_{\algo 4}$}
	\KwFunc{\estimf $(\rr;\cc;t)$}
	\BlankLine
	\KwIn{A point $(\rr;\cc;t)$ in $\spcr \times \spcc \times \reels^+$}
	\BlankLine
	\KwOut{An estimate of $\ff(\rr;\cc;t)$}
	\BlankLine
	\BlankLine
	Sample $T'$ in $(-\infty \intersep t]$, with the density $\pdf{T'}:t' \mapsto \frec [\vecb(t')] \exp \{- \int_{t'}^t \ud t'' \, \frec [\vecb(t'')] \}$: $t'$ is obtained\;
	\If{$t' \leqslant 0$}{
		\Return{$\ff \bm(\rrb(0);\ccb(0);0\bm)$}\;
		\tcp{Ends the recursion}}
	\Else{
		Sample $\CCE$: $\cce$ is obtained\;
		Sample $\UUP$: $\uup$ is obtained\;
		$\ccr \gets \ccb (t') - \cce$\;
		$\ccp \gets \tfrac{1}{2} [ \ccb (t') + \cce + \ccrn \uup ]$\;
		$\ccpe \gets \tfrac{1}{2} [ \ccb (t') + \cce - \ccrn \uup ]$\;
		\BlankLine
		\tcp{Recursively estimate $\ff \bm(\rrb(t');\ccb(t');t'\bm)$}
		 $\Ff_1 \gets \estimf \bm(\rrb(t');\ccb(t');t'\bm)$\;
		\tcp{Recursively estimate $\ff \bm(\rrb(t');\cce;t'\bm)$}  
		$\Ff_2 \gets \estimf \bm(\rrb(t');\cce;t'\bm)$\;
		\tcp{Recursively estimate $\ff \bm(\rrb(t');\ccp;t'\bm)$} 
		$\Ff_3 \gets \estimf \bm(\rrb(t');\ccp;t'\bm)$\;
		\tcp{Recursively estimate $\ff \bm(\rrb(t');\ccpe;t'\bm)$}  
		$\Ff_4 \gets \estimf \bm(\rrb(t');\ccpe;t'\bm)$\;
		\BlankLine
		\Return{$\Ff_1 + \ccrn \sedf (\ccrn;\uup \cdot \uu_\ccr) \, \dfrac{\Ff_3\,\Ff_4 - \Ff_1\,\Ff_2}{\pdf{\CCE}(\cce) \,\pdf{\UUP}(\uup) \,\frec [\vecb(t')]}$}\;
	}
\parbox{\linewidth}{\caption{Estimation of $\ff(\rr;\cc;t)$, valid in the conditions of the harmonic trap described in Secs.~\ref{sec:ht} and \ref{sec:mix}. $\ff$ is supposed to be known at $t = 0$. This algorithm is recursive.}}
\label{algo:ht-simple}
\end{algorithm*}

Algorithm~\ref{algo:ht-simple} has an uncommon feature: if the initial distribution is in the kernel of the collision operator [as described in Eq.~\eqref{eq:ht-moments}], then whatever the choices retained for $\CCE$, $\UUP$ and $T'$ (defined as an exponential RV of parameter $\frec$) and whatever the collision cross-section $\sedf$, the result of Algo.~\ref{algo:ht-simple} has a null variance. To our knowledge, this is the first Monte-Carlo algorithm which exhibits a zero variance result independently of the sampling choices.

One may be confident in the above statement for the following reasons:
\begin{proof}[Null variance of Algo.~\ref{algo:ht-simple}]
Let $\FF (\rr;\cc;t)$ be the estimator of $\ff (\rr;\cc;t)$, obtained through Algo.~\ref{algo:ht-simple}. We assume that the initial distribution $\ff (\rr;\cc;0)$ is everywhere Maxwellian, with the first moments described in Eq.~\eqref{eq:ht-moments}.

A needed lemma here is the following: for any arbitrary choice of $(\rr;\cc;t;t') \in \spcr \times \spcc \times \reels \times \reels$, a value for $\ff(\rr;\cc;t)$ is compatible with a Maxwellian distribution the first moments of which are described in Eq.~\eqref{eq:ht-moments}, if and only if the same value for $\ff\bm(\rrb(t');\ccb(t');t'\bm)$ is compatible with a Maxwellian distribution the first moments of which are described in Eq.~\eqref{eq:ht-moments} with substitution of $t$ by $t'$. This means, by virtue of Liouville's theorem, that an oscillating Maxwellian distribution, as described in Eq.~\eqref{eq:ht-moments}, is compatible with pure ballistic transport in the given force field. The proof of this, and its extension to more general expressions of the collision operator kernel and force field, can be found in \cite{Guery-Odelin2014} and will not be repeated here.

Now consider the following induction hypothesis, to apply anywhere in an estimation tree: $\FF (\rr;\cc;t) = \ff (\rr;\cc;t)$ with null variance, describing a Maxwellian distribution, the first moments of which are given in Eq.~\eqref{eq:ht-moments}.

Starting Algo.~\ref{algo:ht-simple}, there are two possibilities: $T' \leqslant 0$ and a leaf of the estimation tree is reached, or $T' > 0$ and a non-leaf node of the tree is reached.

If $T' \leqslant 0$ then
\begin{equation}
\FF (\rr;\cc;t) = \ff \bm(\rrb(0);\ccb(0);0\bm).
\end{equation}
Because of the initial condition and of the lemma, the induction hypothesis is valid.

If $T' > 0$ then
\begin{align*}
\FF (\rr;\cc;t) &= \FF_1 + \Bedf \,\dfrac{\FF_3 \FF_4 - \FF_1 \FF_2}{\pdf{\CCE}(\CCE) \,\pdf{\UUP}(\UUP) \,\frec [\vecb(T')]} \\[1ex]
&= \FF \bm(\rrb(T');\ccb(T');T'\bm) + \sfo \\[1ex]
&\ \qquad\dfrac{\Bedf}{\pdf{\CCE}(\CCE) \,\pdf{\UUP}(\UUP) \,\frec [\vecb(T')]} \times \sfo \\[1ex]
&\ \qquad \bm[ \FF \bm(\rrb(T');\CCP;T'\bm) \,\FF \bm(\rrb(T');\CCPE;T'\bm) - \sfo \\[1ex]
&\ \qquad \FF \bm(\rrb(T');\ccb(T');T'\bm) \,\FF \bm(\rrb(T');\CCE;T'\bm) \bm],
\end{align*}
where $\FF_1$, $\FF_2$, $\FF_3$, $\FF_4$, $\CCP$, $\CCPE$, and $\Bedf$ are the RVs corresponding to the values $\Ff_1$, $\Ff_2$, $\Ff_3$, $\Ff_4$, $\ccp$, $\ccpe$, and $\ccrn \sedf (\ccrn;\uup\cdot\uu_\ccr)$ in Algo.~\ref{algo:ht-simple}. If the induction hypothesis holds for $\FF_1$, $\FF_2$, $\FF_3$ and $\FF_4$, then
\begin{multline*}
\FF (\rr;\cc;t) = \ff \bm(\rrb(T');\ccb(T');T'\bm) + \sfo \\[1ex]
\dfrac{\Bedf}{\pdf{\CCE}(\CCE) \,\pdf{\UUP}(\UUP) \,\frec \bm(\rrb(T');\ccb(T');T'\bm)} \times \sfo \\[1ex]
\bm[ \ff \bm(\rrb(T');\CCP;T'\bm) \,\ff \bm(\rrb(T');\CCPE;T'\bm) - \sfo \\[1ex]
\ff \bm(\rrb(T');\ccb(T');T'\bm) \,\ff \bm(\rrb(T');\CCE;T'\bm) \bm].
\end{multline*}
Considering now that $\ff$ at $\bm(\rrb(T');T'\bm)$ follows a Max\-wel\-lian distribution, which respects the detailed balance
\begin{equation}
\ff (\rr;\ccp;t) \ff (\rr;\ccpe;t) = \ff (\rr;\cc;t) \ff (\rr;\cce;t),
\end{equation}
one concludes that
\begin{equation}
\FF (\rr;\cc;t) = \ff \bm(\rrb(T');\ccb(T');T'\bm).
\end{equation}
Following the lemma, the induction hypothesis propagates.

As long as the estimation tree is finite (which brings conditions on $\frec$), the induction hypothesis applies from the leaves of the tree to any node of the tree, root included.
\end{proof}

Some numerical experiments (not displayed here) have confirmed this result: if the initial distribution $\ff (\rr;\cc;0)$ belongs to the collision operator kernel, Algo.~\ref{algo:ht-simple} has a null variance.

\section{The harmonic trap\\ without local equilibrium}
\label{sec:mix}

\paragraph{We now test} a case where the initial distribution is outside the collision operator kernel, and follows the BKW mode distribution at maximal disequilibrium, still keeping the same first moments at each position $\rr$. This means that at every point of space the initial distribution of velocities is the one described in Eq.~\eqref{eq:bkw-max-disequilibrium}, translated and scaled to accord with the density, peculiar speed, and temperature prescribed at initial time $t=0$ by Eq.~\eqref{eq:ht-moments}. We have chosen $\rcqm = \frac{9}{41}$ and $\phaz = 0$ (which is also the choice displayed in Fig.~\ref{fig:ht-periodic-breathing} and explained at the end of Par.~\ref{sec:ht-physical situation}; the other parameters $\raid$, $n$, and $\cqmeq$ drop by non-dimensionalization). For the collision model, we have chosen again, for the sake of simplicity, the isotropic Maxwell model; the differential collision cross section equals $\sedf = \kappa / (4 \pi g)$, with $\kappa$ such that $\frac{n \, \kappa \,\raid^2}{\cqmeq^3} = 3$.

After the initial condition, the system evolution will be different from the case in Sec.~\ref{sec:ht}. The differences even appear in macroscopic quantities. As far as the authors know, there is no current explicit description available for this evolution. Nevertheless, two limit situations are easily described:
\begin{itemize}
\item If the cross sections are null (i.e., there are no collisions), the evolution of the system is due only to ballistic transport in the harmonic force field. This is explicitly calculable. 
\item If the cross sections are not null, the final state of the gas is exactly the oscillating state described in the previous section in Eq.~\eqref{eq:ht-moments}~; the collisional invariants in the system ensure that it is the only state belonging to the collision operator kernel that the system can reach.
\end{itemize}

\begin{algorithm*}[!tp]
\newcommand{\intern}[1]{{#1^\diamond}}
	\SetAlgoVlined
	\SetKwFunction{estimf}{$\Ff_{\algo 5}$}
	\KwFunc{\estimf $(\rr;\cc;t)$}
	\BlankLine
	\KwIn{A point $(\rr;\cc;t)$ in $\spcr \times \spcc \times \reels^+$}
	\BlankLine
	\KwOut{An estimate of $\ff(\rr;\cc;t)$}
	\BlankLine
	\BlankLine
	$t' \gets t$\;
	\Repeat{$r\frecc < \frec'$}{
		Sample $\intern{T}$ in $(-\infty \intersep t']$ following an exponential law for $\intern{T} \leqslant t'$ with rate $\frecc$: $\intern{t}$ is obtained\;
		\If{$\intern{t} \leqslant 0$}{
			\Return{$\ff \bm(\rrb(0);\ccb(0);0\bm)$}\;
			\tcp{Ends the recursion}
		}
		Calculate $\cmol_\cok \bm(\rrb(\intern{t});\intern{t}\bm)$; $\intern{\cmol}$ is obtained\;
		$\frec' \gets \frac{1}{e} \bigl( \frac{5}{3} \bigr)\!^\frac{5}{2} \, \kappa \,\intern{\cmol}$\;
		$t' \gets \intern{t}$\;
		Sample $R$ following a uniform law over $[0 \intersep 1]$; $r$ is obtained\;
	}
	Calculate $\cqmcok (t')$; $\cqm'$ is obtained\;
	Calculate $\Mvv_\cok \bm(\rrb(t');t'\bm)$; $\Mvvp$ is obtained\;
	Sample $\CCE$ following a Maxwellian distribution, with peculiar speed $\Mvvp$ and RMS speed $\cqm'$: $\cce$ is obtained\;
	Sample $\UUP$ isotropically: $\uup$ is obtained\;
	$\ccp \gets \tfrac{1}{2} [ \ccb (t') + \cce + \| \ccb (t') - \cce \| \,\uup ]$\;
	$\ccpe \gets \tfrac{1}{2} [ \ccb (t') + \cce - \| \ccb (t') - \cce \| \,\uup ]$\;
	\BlankLine	
	\tcp{Recursively estimate $\ff \bm(\rrb(t');\ccb(t');t'\bm)$}
	$\Ff_1 \gets \estimf \bm(\rrb(t');\ccb(t');t'\bm)$\;
	\tcp{Recursively estimate $\ff \bm(\rrb(t');\cce;t'\bm)$}  
	$\Ff_2 \gets \estimf \bm(\rrb(t');\cce;t'\bm)$\;
	\tcp{Recursively estimate $\ff \bm(\rrb(t');\ccp;t'\bm)$} 
	$\Ff_3 \gets \estimf \bm(\rrb(t');\ccp;t'\bm)$\;
	\tcp{Recursively estimate $\ff \bm(\rrb(t');\ccpe;t'\bm)$}  
	$\Ff_4 \gets \estimf \bm(\rrb(t');\ccpe;t'\bm)$\;
	\BlankLine
	\Return{$\Ff_1 + \dfrac{\kappa \,(\sqrt{2 \pi} \cqm')^3}{\frec'} \exp \Bigl( \dfrac{(\cce - \Mvvp)^2}{2\, \cqm^{\prime\,2}} \Bigr) \,(\Ff_3 \,\Ff_4 - \Ff_1 \,\Ff_2 )$}\;
	
\parbox{\linewidth}{\caption{Estimation of $\ff(\rr;\cc;t)$, valid in the conditions of the harmonic trap described in Secs.~\ref{sec:ht} and \ref{sec:mix}, including the Maxwell collision model. The sampling choices detailed in Sec.~\ref{sec:mix} are used. $\ff$ is supposed to be known at $t = 0$. This algorithm is recursive.}}
\label{algo:ht-detailed}
\end{algorithm*}

\paragraph{To finalize the algorithm,} we need to specify $\UUP$, $\CCE$, and $\frec$. $\UUP$ follows an isotropic law, as in Sec.~\ref{sec:bkw}. $\CCE$ is set to a Maxwellian sampling; the Maxwellian density function is adjusted according to the final distribution, i.e., it is centered on $\Mvv_\cok (\rr;t')$ and scaled to $\cqmcok (t')$ standard deviation given in Eq.~\eqref{eq:ht-moments}, where $t'$ is the collision time. As we wanted $\frec$ as an upper bound of the collision frequency, we have chosen
\begin{equation}
\label{eq:mix-the-raised-frequency}
\frec \equiv \frec (\rr;t) = \dfrac{1}{e} \biggl( \dfrac{5}{3} \biggr)^{\!\raisebox{-1.5ex}{$\frac{5}{2}$}} \,\kappa\, \cmol_\cok (\rr;t),
\end{equation}
where $e = \exp 1$. We thus decided that this upper bound depends on position and time. The prefactor $\frac{1}{e} \bigl( \frac{5}{3} \bigr)\!^\frac{5}{2}$ is the maximal ratio between the BKW mode initial distribution and the equilibrium distribution with the same first moments, i.e.,
\begin{equation}
\dfrac{1}{e} \biggl( \dfrac{5}{3} \biggr)^{\!\raisebox{-1.5ex}{$\frac{5}{2}$}} = \max_{\cc \in \spcc} \, \dfrac{\ff_\indic{BKW} (\cc;0)}{(\sqrt{2 \pi} \cqm)^{-3} \exp [- \ccd / (2 \,\cqm^2)]},
\end{equation}
where $\ff_\indic{BKW} (\cc;0)$ is given by Eq.~\eqref{eq:bkw-max-disequilibrium}.

Sampling a collision time $t'$ using the raised collision frequency $\frec$ given in Eq.~\eqref{eq:mix-the-raised-frequency} can be tricky: it requires one to apply the Beer extinction law to the gas density described in Eq.~\eqref{eq:ht-moments}, along an elliptic ballistic trajectory. To achieve this, we use $\frec$ as the ``true'' extinction frequency in an internal NCA, which in turn uses, as its own raised collision frequency, $\frecc$ defined by
\begin{align}
\label{eq:mix-the-global-raised-frequency}
\frecc &= \dfrac{1}{e} \biggl( \dfrac{5}{3} \biggr)^{\!\raisebox{-1.5ex}{$\frac{5}{2}$}} \,\kappa \,\cmol_\cok (\vnul;t) |_{\sin \phas (t) = -1} \\
 &= \dfrac{1}{e} \biggl( \dfrac{5}{3} \biggr)^{\!\raisebox{-1.5ex}{$\frac{5}{2}$}} \,\kappa \,n \,\biggl( \dfrac{\raid}{\sqrt{2 \pi} \sqrt{1 - \rcqm} \,\cqmeq} \biggr)^{\!\raisebox{-.5ex}{$\scriptstyle 3$}}.
\end{align}
It is the global maximum of $\frec$ defined in Eq.~\eqref{eq:mix-the-raised-frequency}, and is reached at the center of the gas cloud at its maximum contraction.
With all these choices, Algo.~\ref{algo:ht-simple} becomes Algo.~\ref{algo:ht-detailed}.

\begin{figure*}
\parbox{\Graphwidth}{\includegraphics{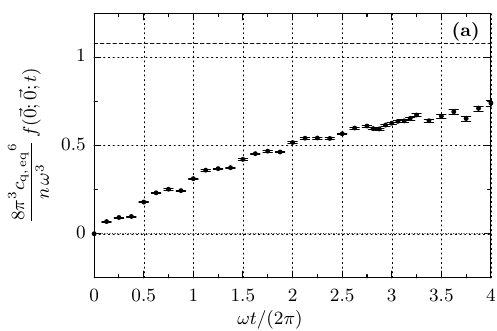}
\hfill
\includegraphics{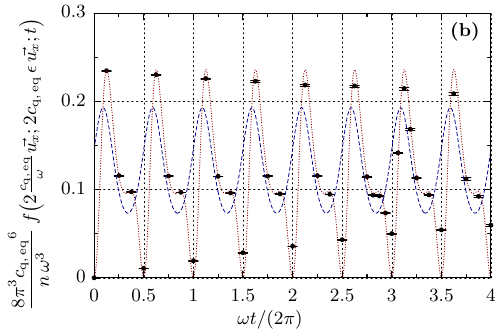}}
\\ \vspace{\baselineskip}
\parbox{\Graphwidth}{\includegraphics{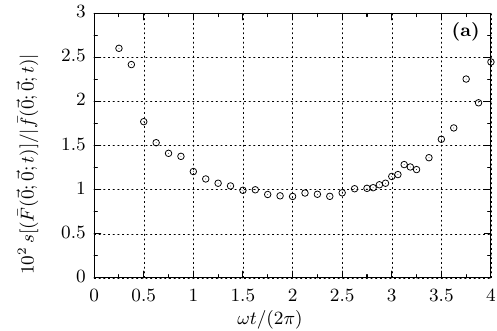}
\hfill
\includegraphics{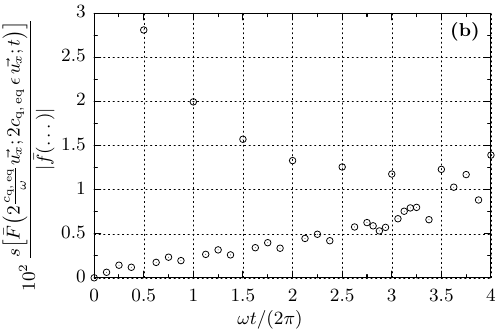}}
\\ \vspace{\baselineskip}
\parbox{\Graphwidth}{\includegraphics{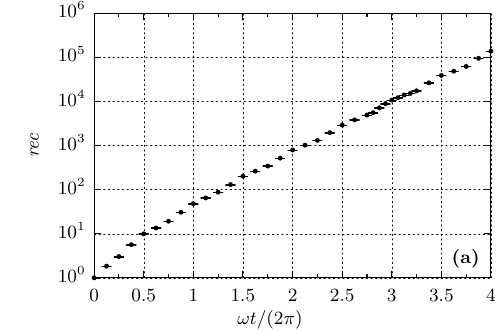}
\hfill
\includegraphics{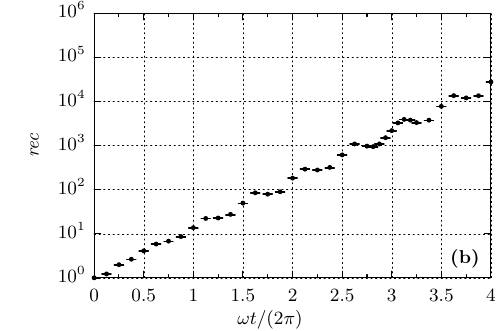}}
\\
\caption{Results obtained by Algo.~\ref{algo:ht-detailed} applied to the harmonic trap problem described in Secs.~\ref{sec:ht} and \ref{sec:mix}. We follow the distribution function $\ff$ along the time $t$, at two probe points: the center of the phase space in column~(a), and a point located $2\times$ the RMS radius away the origin in column~(b). These probe points $(\rr;\cc)$ comply with the constraint $\cc = \Mvv_\cok (\rr;0)$, such that there $\ff = 0$ at the initial instant.\\
The confidence intervals displayed in the top and bottom graphs take account of 1 standard deviation. The graphs at the top give a graphical comparison between the results, the values predicted in the final oscillating state (blue dashed lines), and the values predicted by ballistic transport of the initial condition (red dotted lines). The graphs in the middle line show, with the same results, the relative standard deviation given by the calculation. The graphs at the bottom display the mean recursivity of Algo.~\ref{algo:ht-detailed} computing the points on the graphs above. Each displayed point has been obtained through running \num{e4} realizations of Algo.~\ref{algo:ht-detailed}.\\
In every graph, each point has been calculated independently [points are of course the same at each abscissa along column (a) and along column (b)]. This is how we have, naturally, placed more probe dates around $\raid t/(2\pi) = 3$.}
\label{fig:mix-res-sonde}
\end{figure*}

\begin{figure*}
\parbox{\Graphwidth}{\includegraphics{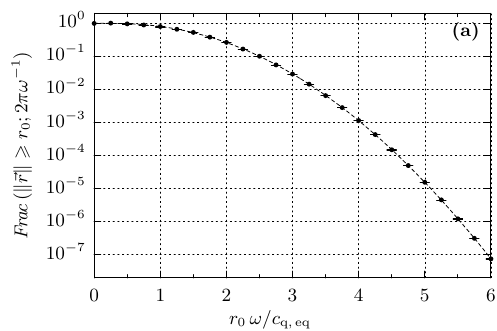}
\hfill
\includegraphics{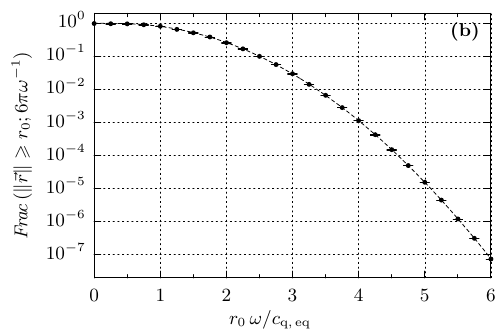}}
\\ \vspace{\baselineskip}
\parbox{\Graphwidth}{\includegraphics{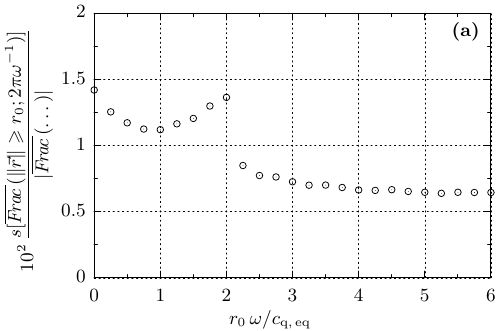}
\hfill
\includegraphics{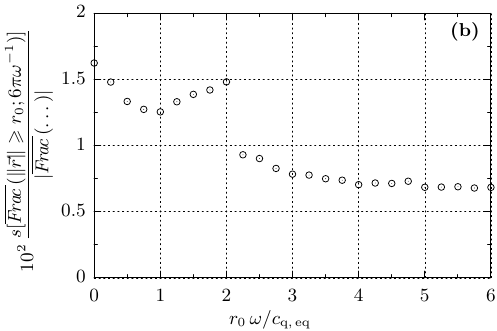}}
\\ \vspace{\baselineskip}
\parbox{\Graphwidth}{\includegraphics{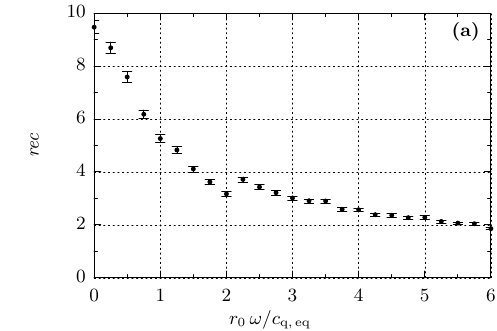}
\hfill
\includegraphics{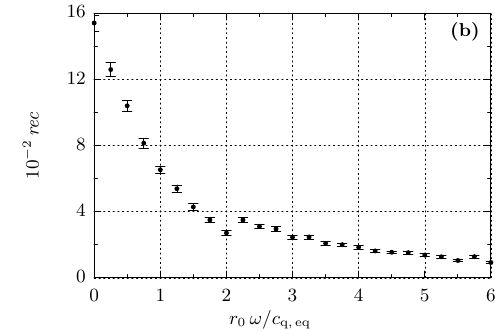}}
\\
\caption{Results obtained by Algo.~\ref{algo:HR-integrator} combined with the sampling laws listed in Eq.~\eqref{eq:HR-sampling}, and applied to the harmonic trap problem described in Secs.~\ref{sec:ht} and \ref{sec:mix}. What was calculated is the distribution of particles along the radius $\|\rr\|$, i.e., how many particles are more distant from the origin than any given threshold $r_0$; two dates are considered, $\raid t = 2\pi$ in column~(a), and $\raid t = 6\pi$ in column~(b).\\
The confidence intervals displayed in the top and bottom graphs take account of 1 standard deviation. The graphs at the top give a graphical comparison between the results and the values predicted in the final oscillating state (dashed lines; ballistic transport of the initial condition leads to the same results, in these precise cases). The graphs in the middle line show, with the same results, the relative standard deviation given by the calculation. The graphs at the bottom display the mean recursivity of Algo.~\ref{algo:ht-detailed} called by Algo.~\ref{algo:HR-integrator}, when computing the points on the graphs above. Each displayed point was obtained through running \num{e4} realizations of Algo.~\ref{algo:HR-integrator}.}
\label{fig:mix-res-GR}
\end{figure*}

\paragraph{Algorithm~\ref{algo:ht-detailed} was actually run,}  and results are displayed in Figs.~\ref{fig:mix-res-sonde} and \ref{fig:mix-res-GR}. In Fig.~\ref{fig:mix-res-sonde} the distribution function $\ff$ is followed at two probe points of the phase space: at the center of the phase space, and at another point placed 2× the RMS radius away from the center of positions. %
Figure~\ref{fig:mix-res-GR} shows the tail distribution of the mass along the potential energy (or equivalently, along the distance from the origin). This tail is calculated using Algo.~\ref{algo:HR-integrator}, which encapsulates Algo.~\ref{algo:ht-detailed} with the additional sampling of a final point $(\RR_\final;\CC_\final)$ of the particle tree, more distant to the origin than a given value $r_0$. The sampling law of $(\RR_\final;\CC_\final)$ in Algo.~\ref{algo:HR-integrator} is similar to the one given in Eq.~\eqref{eq:HS-sampling}, with
\begin{equation}
\label{eq:HR-sampling}
\left\{\begin{array}{@{\,}l@{}}
\RR_\final = R_\final \,\UU_\final \quad \texte{with} \\[1ex]
\left\{\begin{array}{@{\,}l@{}}
\pdf{R_\final} (r_\final) = \sfo \\
\ \left\{\begin{array}{@{\,}l@{\ }l@{}}
\dfrac{r_0 \Bigl[ \Bigl(\dfrac{\raid\, r_0}{\cqmeq}\Bigr)^{\!2} - 2\Bigr]}{\Bigl\{ \Bigl[ \Bigl(\dfrac{\raid\, r_0}{\cqmeq}\Bigr)^{\!2} - 2 \Bigr] (r_\final - r_0) + r_0 \Bigr\}\big.^{\!2}} & \texte{if } r_0 > \dfrac{2\,\cqmeq}{\raid} \\ \\[-1ex]
\dfrac{2 \cqmeq / \raid}{(2 \cqmeq / \raid + r_\final - r_0)^2} & \texte{if } r_0 \leqslant \dfrac{2\,\cqmeq}{\raid} \\
\end{array}\right. \\ \\[-1ex]
\UU_\final \texte{ of isotropic law} \\
\end{array}\right. \\ \\[-1ex]
\CC_\final \texte{ of law } \doms{N} \bm( \Mvv_\cok (\RR_\final;t); \cqmcok(t)^2 \matid \bm) .
\end{array}\right.
\end{equation}

\begin{algorithm}[!ht]
	\SetAlgoVlined
	\KwIn{A radius $r_0$ and a time $t$}
	\BlankLine
	\KwOut{An estimate of $\Frac(\| \rr \| \geqslant r_0; t)$, the fraction of particles more distant than $r_0$ from the origin at time $t$}
	\BlankLine
	Sample $R_\final$: $r_\final$ is obtained\tcp*[r]{$r_\final \geqslant r_0$}
	Sample $\UU_\final$: $\uu_\final$ is obtained\;
	Sample $\CC_\final$: $\cc_\final$ is obtained\;
	Estimate $\ff (r_\final \mspace{1mu} \uu_\final; \cc_\final; t)$ using Algo.~\ref{algo:ht-detailed}: $\Ff$ is obtained\;
	\Return{$\dfrac{r_\final\sfo^2 \, \Ff}{\pdf{R_\final}(r_\final) \, \pdf{\UU_\final}(\uu_\final) \, \pdf{\CC_\final}(\cc_\final)}$}\;
\parbox{\linewidth}{\caption{Estimation of a fraction of particles with high potential energy, valid in the conditions of the harmonic trap described in Secs.~\ref{sec:ht} and \ref{sec:mix}}}
\label{algo:HR-integrator}
\end{algorithm}

We observe the same behaviors in Algos.~\ref{algo:ht-detailed} and \ref{algo:HR-integrator} as in Algos.~\ref{algo:bkw-simple}-bkw and \ref{algo:HS-integrator}:

The variance of the obtained estimator of $\ff$ increases with the simulated physical time, as shown in particular in Fig.~\ref{fig:mix-res-sonde}. The middle line of the left panel shows a dynamic having a minimum with time, but what is displayed is a \emph{relative} variance. The increase of this relative variance near $t=0$ is due only to the fact that the estimated quantity $\ff (\vnul;\vnul;t)$ starts at zero when $t=0$; it shows that when $t \to 0$, $\ff (\vnul;\vnul;t)$ tends to zero faster than the standard deviation of its estimator. The same dynamic is visible on the relative variance exposed in the left panels in Fig.~\ref{fig:bkw-res-pos-fixe}. The maximums observed in the relative variance exposed in the middle line panel of Fig.~\ref{fig:mix-res-sonde}(b) can be explained the same way: they correspond to minimums of the estimated quantity which are reached periodically, as shown in the panel above. See, by the way, how these maximums of the relative variance decrease with time, as the minimums of $\ff$ move away from 0.

The relative variance of the results of Algos.~\ref{algo:ht-detailed} and \ref{algo:HR-integrator} is practically insensitive to rarefaction; we can evaluate without difficulty fractions of the mass as small as one millionth.

There is a noticeable decrease in the recursivity of Algo.~\ref{algo:ht-detailed} when the probe points are taken farther from the origin. This is expected since the raised collision frequency $\frec$ is higher near the origin. The recursion depth has been diminished by the choice of space-time-dependent $\frec$ described in Eq.~\eqref{eq:mix-the-raised-frequency}. If we had used instead a constant raised collision frequency such as $\frecc$ described in Eq.~\eqref{eq:mix-the-global-raised-frequency}, the mean recursion would have been higher. 

The discontinuities visible on the bottom panels of Fig.~\ref{fig:mix-res-GR} at $\raid \,r_0 = 2 \,\cqmeq$, are due to the discontinuous change of the sampling law of $R_\final$ as listed in Eq.~\eqref{eq:HR-sampling}. The same effect can be seen on the Fig.~\ref{fig:bkw-res-Qft} at $c_0 = 2 \,\cqm$. These different sampling laws are designed to keep a low variance of the estimations.
The change of sampling law does not bias the obtained estimators, as they are finally divided by the sampling probability densities to keep the wanted expectations, as explained in Eq.~\eqref{eq:intro-MMC-principle}; but it does change their variance.
In cases displayed in Fig.~\ref{fig:mix-res-GR}, it also changes slightly the mean recursion of called Algo.~\ref{algo:ht-detailed}: as explained just above, this mean recursion decreases with the distance to the origin, and precisely this distance is distributed differently when $r_0$ changes.

\section{Conclusion and perspectives}
\label{sec:persp}

\paragraph{} We have numerically solved the Boltzmann equation of gas kinetics in several test cases, using the Monte-Carlo method. It is the Monte-Carlo method as used in linear transport physics: without bias, and where particle paths are computed independently. The algorithms that we have proposed are based on an integral formulation of the Boltzmann equation, and rely on the following of virtual particles (and their collision partners) from their arrivals to their sources. As far as the authors know, this is the first numerical method in gas kinetics which thus proceeds along time, backwards.

The method retains some advantages of the MCM from linear transport physics, its most striking properties being the following:
\begin{itemize}
\item Parallelization of the algorithm remains trivial.
\item No mesh or time discretization is necessary in the method.
\item The method enables probe calculation, i.e., it is possible to calculate the distribution function at a point of the phase space without computing the rest of the field.
\item Rarefaction of the gas at the probe does not compromise the relative accuracy of the calculations.
\item Similarly, the frequency of rare events (in the space of speeds) can be estimated however scarce they may be.
\item The method is limited in (spatial or temporal) Knudsen number; because it uses a branching estimation process, the mean complexity of which grows exponentially with the mean number of collisions.
\item The branchings in the estimation process also increase the variance of the obtained estimator at low Knudsen, as briefly explained in Par.~\ref{sec:bkw-result-comments}. Several ideas could be explored to lower this variance:
\begin{itemize}
\item Following the proposition in \cite{Weitz2016}, one could evaluate several times the quantity of collision partners at each collision. This will be the subject of future work.
\item The control variate technique could be used to lower the variance produced by each sampling in the estimation process. The control variables could be built on an approximative solution of the problem under study, such as the BGK phenomenology \cite{Bhatnagar1954}. This will also be the subject of future work.
\item The estimation process could be replaced, beyond a given branching depth, by an approximative solution. This amounts to sacrifice the exactitude of our Monte-Carlo method to avoid a high branching recursion and the aforementioned associated problems. Similar ideas have been explored in \cite{Tregan2020a}.
\end{itemize}
\item As a result of the two previous points, we are today essentially unable to perform a calculation on a steady state.
\end{itemize}

\paragraph{If a geometry is present}\label{sec:persp-geometrical-complexity} (a possibility not illustrated in this article), it enters the integral formulation of the Boltzmann equation only through a calculation of intersections between surfaces and ballistic trajectories. In addition, the boundary conditions in gas kinetics are generally written as integral expressions of the distribution of the outgoing molecules, directly usable in our Monte-Carlo method. This leads us to state that taking into account any geometry will bring no difficulty, given that ballistic trajectories are simple enough (typically, straight lines). If very complex geometries come into play, we will be able to use the expertise of the image synthesis community directly \cite{DelaTorre2014}.

\paragraph{} Because the presented Monte-Carlo method basically performs probe estimations, benchmark comparison with other algorithms which compute the whole density field at once (such as Direct Simulation Monte Carlo methods \cite{Bird1994, Bird1998, Oran1998, Khisamutdinov2004, Homolle2007, Degond2010, Stefanov2011, Murrone2011}, Lattice Boltzmann methods \cite{Piaud2014, Ambrus2016, Sofonea2003, He1997a, He1997b, Chen1998}, Unified Gas Kinetic Schemes \cite{Chen2012a, Huang2012, Guo2013}, Fast Kinetic Schemes \cite{Dimarco2013, Dimarco2013b, Dimarco2015}, or Discrete Velocity Methods \cite{Sharipov1993, Palczewski1997, Palczewski1998, Mieussens2000}\dots) would be irrelevant.

Here are four points we believe suitable for comparing our method to the other ones:

\begin{itemize}

\item The main drawback of our method is that it works only in physical situations with high Knudsen number.
For example, in the physical situations presented in Secs.~\ref{sec:bkw} and \ref{sec:mix}, it allows probe calculations only a few free flight mean times after the initial condition.
Other numerical methods do not suffer from such a constraint.

\item Even though mathematical developments shown in Pars.~\ref{sec:algo-2-sampling-and-properties} and \ref{sec:ht-simple-algo} indicate that our method performs better (i.e., it gives estimators with lower variance) in gas near equilibrium, our method deals well with gas in sharp disequilibrium, as illustrated in Secs.~\ref{sec:bkw} and \ref{sec:mix}.
In these situations, our method stays exact up to a reasonable statistical noise.

\item Our method uses the original Boltzmann collision operator, while the aforementioned methods which use discretized velocity space (LBM, UGKS, FKS, DVM) frequently rely on the approximate BGK collision operator.

\item Our method is also able to probe the quantity of molecules in unpopulated parts of the phase space, for example, at high kinetic energy, however scarce they may be, without loss of relative accuracy. The tests presented in Secs.~\ref{sec:bkw} and \ref{sec:mix} show calculations which would be unaffordable with any other method.

\end{itemize}

\begin{acknowledgments}
This work has been sponsored by the French government research program ``Investissements d'avenir'' through the ANR program ANR-12-IS04-0003 \mbox{DEPART}, the Laboratories of Excellence ProjetIA-10-LABX-0022 \mbox{SOLSTICE} and ANR-10-LABX-0016 \mbox{IMobS\textsuperscript{3}}, the ATS program \mbox{ALGUE} of the IDEX of Toulouse ANR-11-IDEX-0002 \mbox{UNITI}, by the Occitanie region, and by the \mbox{LAPLACE} laboratory through a BQR program.
\end{acknowledgments}



\label{DocEnd}
\end{document}